\documentclass[11pt]{article}
\pdfoutput=1
\usepackage{jheppub}
\usepackage{enumerate}
\usepackage{shuffle}
\usepackage{tikz}
\usetikzlibrary{calc,decorations.pathreplacing}

\usepackage{hyperref}
\hypersetup{colorlinks=true,linkcolor=magenta,anchorcolor=green,citecolor=cyan,filecolor=black,menucolor=black,urlcolor=blue}

\def\Ree{\Re\textrm{e}}
\def\Imm{\Im\textrm{m}}
\def\tr(#1){\textrm{tr}\left(#1\right)}
\def\trF(#1){\textrm{tr}_{\rm fund}\left(#1\right)}
\def\trA(#1){\textrm{tr}_{\rm adj}\left(#1\right)}

\def\ket#1{|#1 \rangle}

\def\la{\lambda}
\def\lb{\tilde{\lambda}}

\def\Atree{A^{\rm tree}}
\def\cAtree{\mathcal A^{\rm tree}}
\def\hAtree{\hat A^{\rm tree}}
\def\Acut{A^{1\textrm{-cut}}}
\def\Aone{A^{(1)}}

\def\cA{\mathcal A}

\def\cS{\mathcal S}

\def\mS{\mathfrak S}

\def\dd{d\!\!{}^-\!}

\def\ps(#1){p_{\sigma(#1)}}

\def\c(\{p_#1,p_#2\},p_#3,p_#4,p_#5){c_{|#1#2|#3#4#5}}

\def\be{\begin{equation}}
\def\ee{\end{equation}}
\def\bea{\begin{eqnarray}}
\def\eea{\end{eqnarray}}
\def\beal{\begin{equation}\begin{aligned}}
\def\eeal{\end{aligned}\end{equation}}
\def\nn{\nonumber}
\def\eqn#1{eq.~\eqref{#1}}
\def\eqns#1#2{eqs.~\eqref{#1} and~\eqref{#2}}
\def\eqnss#1#2#3{eqs.~\eqref{#1}, \eqref{#2} and~\eqref{#3}}
\def\Eqn#1{Eq.~\eqref{#1}}
\def\Eqns#1#2{Eqs.~\eqref{#1} and~\eqref{#2}}
\def\fig#1{figure~{\ref{#1}}}
\def\figs#1#2{figures~{\ref{#1}} and~{\ref{#2}}}
\def\Fig#1{Figure~{\ref{#1}}}
\def\Figs#1#2{Figures~{\ref{#1}} and~{\ref{#2}}}
\def\tab#1{table~{\ref{#1}}}
\def\tabs#1#2{tables~{\ref{#1}} and~{\ref{#2}}}
\def\Tab#1{Table~{\ref{#1}}}
\def\Tabs#1#2{Tables~{\ref{#1}} and~{\ref{#2}}}
\def\sec#1{section~{\ref{#1}}}
\def\secs#1#2{section~{\ref{#1}} and~{\ref{#2}}}
\def\Sec#1{Section~{\ref{#1}}}
\def\Secs#1#2{Section~{\ref{#1}} and~{\ref{#2}}}
\def\app#1{appendix~{\ref{#1}}}
\def\draftnote#1{{\color{red} #1}}
\newcommand{\usegraph}[1]{\vcenter{\hbox{\!\;\includegraphics[scale=1.0]{graphs/#1.pdf}\!\;}}}
\newcommand{\scalegraph}[2]{\vcenter{\hbox{\!\;\includegraphics[scale=#1]{graphs/#2.pdf}\!\;}}}

\newcount\hh
\newcount\mm
\mm=\time
\hh=\time
\divide\hh by 60
\divide\mm by 60
\multiply\mm by 60
\mm=-\mm
\advance\mm by \time
\def\thistime{\number\hh:\ifnum\mm<10{}0\fi\number\mm}

\begin{document}
\preprint{DAMTP-2017-31, Edinburgh 2017/16, IPhT-t17/065\\}

\title{One-loop monodromy relations on single cuts}

\author[1]{Alexander Ochirov,} \author[2]{Piotr Tourkine}
\author[3]{\textrm{and} Pierre Vanhove}
\affiliation[1]{Higgs Centre for Theoretical Physics, School of Physics and Astronomy,\\
  The University of Edinburgh, Peter Guthrie Tait Road, Edinburgh EH9
  3FD, Scotland, UK}
\affiliation[2]{Department of Applied Mathematics and Theoretical Physics
  (DAMTP),\\
  University of Cambridge, Wilberforce Road, Cambridge CB3 0WA, UK}
\affiliation[3]{Institut de Physique Th\'eorique,\\
  Universit\'e Paris Saclay, CNRS, F-91191 Gif-sur-Yvette, France}
\emailAdd{alexander.ochirov@ed.ac.uk, pt373@cam.ac.uk,
  pierre.vanhove@ipht.fr} 
\abstract{The discovery of colour-kinematic duality has led to significant progress in the computation of scattering amplitudes in quantum field theories. At tree level, the origin of the duality can be traced back to the monodromies of open-string amplitudes. This construction has recently been extended to all loop orders. In the present paper, we dissect some consequences of these new monodromy relations at one loop. We use single cuts in order to relate them to the tree-level relations. We show that there are new classes of kinematically independent single-cut amplitudes. Then we turn to the Feynman diagrammatics of the string-theory monodromy relations. We revisit the string-theoretic derivation and argue that some terms, that vanish upon integration in string and field theory, provide a characterisation of momentum-shifting ambiguities in these representations. We observe that colour-dual representations are compatible with this analysis.
}
\maketitle
\newpage

\section{Introduction}
\label{sec:introduction} 

It is truly surprising that,
more than sixty years after the basic principles of
quantum field theory have been spelled out,
we continue to discover profound facts about its perturbative expansion.
The discovery of colour-kinematic duality~\cite{Bern:2008qj,Bern:2010ue},
for instance, provided a new way to organise the perturbation theory,
which reveals a network of relationships
between a variety of gauge, gravity and scalar field theories.

As a consequence of this duality, the number of independent tree-level
colour-ordered amplitudes for $n$ gluons turned out to be $(n-3)!$, whereas
colour structure alone gives $(n-2)!$~\cite{Kleiss:1988ne}. This
aspect of the duality is beautifully explained via string theory by
the monodromy properties of open-string
amplitudes~\cite{BjerrumBohr:2009rd,BjerrumBohr:2010zs,Stieberger:2009hq}
and also in a Feynman-diagram context~\cite{Mafra:2011kj}.
The duality holds in various
settings~\cite{Monteiro:2011pc,Chiodaroli:2013upa,Johansson:2014zca,Chiodaroli:2014xia,He:2015wgf,Johansson:2015oia,Mogull:2015adi,Johansson:2017bfl,Chiodaroli:2017ngp}
and up to the fourth loop
order~\cite{Bern:2013qca,Bern:2014sna}. However, its extension to all
orders~\cite{Bern:2010ue} remains conjectural, and a modified implementation including contact terms has been
proposed~\cite{Bern:2017yxu} to tackle the five-loop question, similar
in spirit to the construction of~\cite{BjerrumBohr:2010zs}.

In~\cite{Tourkine:2016bak} two of the present authors
have generalised the open-string monodromy relations to the multi-loop case
and proposed corresponding explicit formulae.
That construction generalised some
relations for one-loop integrands made earlier
in~\cite{Boels:2011tp,Boels:2011mn,Feng:2011fja,Brown:2016hck,Brown:2016mrh}.
Implications for the colour-kinematic duality itself were left open: the
purpose of this paper is to study some of these at the one-loop level.

We start by formulating in \sec{sec:single-cut}
a different formulation of the monodromy relations,
based on single cuts of field and string theory amplitudes.
This leads us to some relations between regularised
forward tree-amplitudes, somehow similar to
these of~\cite{He:2016mzd,He:2017spx} based on the Cachazo, He and
Yuan formalism (CHY)~\cite{Cachazo:2013hca,Cachazo:2013iea}. 
These relations involve the forward limit
of the so-called momentum kernel~\cite{BjerrumBohr:2010hn}.
The equivalence between this formulation and the one
of~\cite{Tourkine:2016bak} is shown later on in \sec{sec:bcj-comp-monodr}.
Our main results are the following:
\begin{itemize}
\item We compute the co-rank of the momentum kernel in the forward
  limit: it increases from $(n-1)!$ to $(n-1)!+(n-2)!+2(n-3)!$
  independent regulated $(n+2)$-point forward tree amplitudes.  We
  identify three classes of independent tree amplitudes in the forward
  limit. These classes play a role in amplitudes where non-planar
  cuts are needed.

\item We derive the forward monodromy relations from the monodromies
  in string theory. We re-interpret these monodromies, which actually
  produce {\emph exact} integrand relations in the field-theory
  limit, by providing a precise form for the ambiguities that vanish
  upon integration in field theory. We explain the connection between
  such relations and the Bern-Carrasco-Johansson (BCJ) identities for
  forward tree amplitudes. 
\item We further show on a few examples how these integrand relations are
   always satisfied if a colour-dual representation\footnote{
Colour-kinematic duality~\cite{Bern:2008qj,Bern:2010ue}
requires that the kinematic numerators of a gauge-theory amplitude
satisfy (or can be massaged to satisfy)
the same algebraic relations as their corresponding colour factors,
for example, kinematic Jacobi identities.
The duality-satisfying amplitude representation is then called colour-dual.
}
   of the loop amplitude exists. This is bootstrapped up to
   a consistency constraint on the form a string-theoretic integrand that
   would produce colour-dual kinematic numerators. We observe the
   natural appearance of higher-order Jacobi-type relations.
\end{itemize}

We also present two appendices.  In~\app{sec:forward-limit-param} we
give details on the way to perform the forward limit.
In~\app{sec:one-loop-monodromies}, we check that the monodromy
relations hold to the first order in $\alpha'$ for open-string at one loop.
At the same time we solve the apparent discrepancy between 
the results of~\cite{Tourkine:2016bak} and \cite{Hohenegger:2017kqy}
by showing that it amounts to the different definitions of the branch cuts of the complex logarithms. Once this is taken into account, agreement is obtained.

\section{Single cut of one-loop amplitudes in field and string theory}
\label{sec:single-cut}

\subsection{Single cut and forward tree amplitudes}
\label{sec:part-integr-single}

In this section we discuss the reconstruction of one-loop amplitudes
from single cuts.
In the process we set up our conventions for the rest of the paper.

In any local unitary quantum field theory,
for any number of spacetime dimensions,
the Feynman Tree theorem~\cite{Feynman:1963ax,Feynman:1972mt}
expresses one-loop amplitudes as integrals of tree-level amplitudes.
The theorem prescribes to sum over all possible ways of cutting an internal
propagator.
Its application to construct one-loop amplitudes~\cite{Brandhuber:2005kd}
can be aided by using MHV vertices~\cite{Cachazo:2004kj}.
In~\cite{Catani:2008xa} it was further shown that a one-loop amplitude
can be obtained by integration of the single-cut amplitude 
over a special contour. This amounts to a change of the $i\varepsilon$-prescription in the propagator,
which is equivalent to the Feynman Tree theorem
and is the basis of the ``loop-tree'' duality method~\cite{Catani:2008xa,
Bierenbaum:2010cy,Bierenbaum:2011gg,Bierenbaum:2012th,Buchta:2014dfa}.

Moreover, in~\cite{NigelGlover:2008ur,Britto:2010um,Britto:2011cr}
it was shown how a QCD amplitude can be calculated by taking single cuts
in $D=4-2\epsilon$.
We follow the logic of these works for relating one-loop amplitudes
and their single cuts.

Consider a colour-stripped $n$-point amplitude in $D$ dimensions,
\begin{equation}
   \Aone_n(p_1,\dots,p_n) = \int\!\dd^{D}\ell\,
      {\mathcal N(\ell) \over \prod_{i=0}^{n-1} d_i} ,
   \qquad
   d_i= (\ell+p_{1\ldots i})^2-m_i^2+i\varepsilon,
\end{equation}
with $\varepsilon>0$, $p_{1 \ldots k} := p_1+\dots+p_k$, and the momentum conservation condition is
$p_{1 \ldots n} =0$. The loop measure is defined by
\begin{equation}
\int\!\dd^{D}\ell := \int\!\frac{d^{D}\ell}{(2\pi)^D} .
\label{eq:loop-measure}
\end{equation}
The single-cut operator $\delta_{d_i}^P$ acts by replacing, in the
integrand of the one-loop integral, the propagator $d_i$ between the
legs $p_i$ and $p_{i+1}$ by the operator
$2\pi\delta^{(+)}((\ell+p_{1\ldots
  i})^2-m_i^2)=2\pi\delta((\ell+p_{1\ldots
  i})^2-m_i^2)\theta((\ell+p_{1\ldots i})^0)$, so that
\begin{equation}
   \delta_{d_i}^P \Aone_n(p_1,\dots,p_n) :=
   \int \dd^{D}\ell~~
   2\pi \delta^{(+)}(d_i)
   \Acut_{n+2}(\ell+p_{1\ldots i},p_{i+1},\dots,p_{n+i},
             -(\ell+p_{1\ldots i})) ,
\end{equation}
where the single-cut amplitude reads
\begin{equation}
  \Acut_{n+2}(\ell+p_{1\ldots i},p_{i+1},\dots,p_{n+i},-(\ell+p_{1\ldots i}))
    = {\mathcal{N}(\ell)\over d_0 d_1\cdots d_{i-1}d_{i+1}\cdots d_{n-1}} .
\end{equation}
By shifting the loop momentum we rewrite the cut propagator as
$d_i=\ell^2-m_i^2$ and the single-cut operator of one-loop integral reads
\begin{equation}\label{eq:CutDef}
   \delta_{d_i}^P \Aone_n(p_1,\dots,p_n) =
   \int \dd^{D}\ell ~~
   2\pi   \delta^{(+)}(\ell^2-m_{\rho(i)}^2)\,
      \Acut_{n+2} (\ell,p_{i+1},\dots,p_{n+i},-\ell) ,
\end{equation}
where the external-particle labels are understood modulo $n$.

The one-loop amplitude is also expressible
in terms of irreducible numerators~\cite{Ossola:2006us}:
\begin{equation}
   \Aone_n(p_1,\dots,p_n)
    = \sum_r \sum_{i_1 < \dots < i_r} \int\!\dd^{D}\ell\,
      {\Delta_{i_1,\dots,i_r}(\ell)\over d_{i_1}\!\dots d_{i_r}} ,
\end{equation}
where the index $r$ runs over all possible irreducible topologies.
For instance, in $D=4-2\epsilon$ we have $r=5$ for pentagons,
$r=4$ for boxes, $r=3$ for triangles, $r=2$ for bubbles
and $r=1$ for tapdole integral functions~\cite{Bern:1994zx,Bern:1994cg}.
Each irreducible numerator $\Delta_{i_1,\dots,i_r}$
is obtained by computing residues on the locus
of vanishing of its corresponding $r$ propagators,
$d_{i_1}=\dots=d_{i_r}=0$~\cite{Britto:2004nc,Forde:2007mi}. 
There are up to $\frac{n!}{r!(n-r)!}$ massive $r$-gon integral
functions in an $n$-point amplitude.

The single cut of a one-loop amplitude with 
propagator $d_i$ removed then decomposes as
\be
  \Acut_{n+2}(\ell+p_{1\ldots i},p_{i+1},\dots,p_{n+i},-(\ell+p_{1\ldots i}))
    = \sum_r \sum_{i< i_1 < \dots < i_{r-1}}\!\!
      \frac{\Delta_{i,i_1,\dots,i_{r-1}}(\ell)}
           {d_{i_1}\!\dots d_{i_{r-1}}} .
\ee
In other words, only the master integrals that have a
propagator between the legs $p_i$ and $p_{i+1}$ contribute to the single cut.
Such a single-cut amplitude can be further cut to gradually
obtain all its irreducible numerators.
In this sense the knowledge of the single cuts of a one-loop amplitude
allows to reconstruct it in full~\cite{NigelGlover:2008ur,Britto:2010um,Britto:2011cr}.

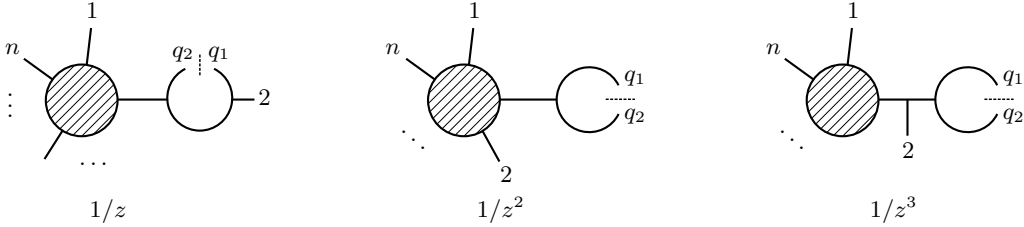
\begin{figure}[t]
   \centering
\footnotesize
\definecolor{cffffff}{RGB}{255,255,255}

\begin{tikzpicture}[y=0.80pt, x=0.80pt, yscale=-1.000000, xscale=1.000000, inner sep=0pt, outer sep=0pt]
  \path[draw=black,line join=miter,line cap=butt,even odd rule,line width=0.800pt]
    (248.0409,34.8055) -- (244.0694,68.3988) --
    (260.3959,97.9765)(216.2967,49.3343) -- (242.9824,68.6580) --
    (292.3817,68.6582);
  \path[draw=black,fill=cffffff,line join=miter,line cap=butt,miter
    limit=4.00,nonzero rule,line width=0.800pt] (243.6106,69.0493) circle
    (0.4768cm);
  \path[draw=black,fill=cffffff,miter limit=4.00,line width=0.320pt]
    (240.1687,52.5460) -- (227.1087,65.6059)(226.8150,70.5131) --
    (245.0766,52.2515)(248.8986,53.0430) -- (227.6035,74.3381)(229.0123,77.5427)
    -- (252.1039,54.4511)(254.7612,56.4073) --
    (230.9681,80.2003)(233.3305,82.4514) -- (257.0134,58.7685)(258.7535,61.6418)
    -- (236.2045,84.1909)(239.5735,85.4354) --
    (259.9937,65.0151)(260.4985,69.1238) -- (243.6850,85.9373)(249.3234,84.9124)
    -- (259.4726,74.7632);
  \path[fill=black,line join=miter,line cap=butt,line width=0.800pt]
    (260.8211,107.8177) node[above right] (text6759) {$2$};
  \path[fill=black,line join=miter,line cap=butt,line width=0.800pt]
    (245.7826,30.2179) node[above right] (text6775) {$1$};
  \path[fill=black,line join=miter,line cap=butt,line width=0.800pt]
    (319.4291,62.0211) node[above right] (text6109-7) {$q_1$};
  \path[cm={{0.0,1.0,-1.0,0.0,(0.0,0.0)}},draw=black,fill=cffffff,miter
    limit=4.00,nonzero rule,line width=0.800pt]
    (75.4796,-316.7471)arc(-63.858:13.004:15.350826 and
    15.424)arc(13.004:89.866:15.350826 and 15.424)arc(89.866:166.728:15.350826 and
    15.424)arc(166.728:243.590:15.350826 and 15.424);
  \path[fill=black,line join=miter,line cap=butt,line width=0.800pt]
    (319.4291,80.1973) node[above right] (text8667) {$q_2$};
  \path[fill=black,line join=miter,line cap=butt,line width=0.800pt]
    (213.0193,93.2483) node[above right] (text6767-4) {$\ddots$};
  \path[draw=black,dash pattern=on 1.12pt off 0.56pt,line join=miter,line
    cap=butt,miter limit=4.00,even odd rule,line width=0.560pt] (310.6705,68.5220)
    -- (324.4697,68.5220);
  \path[fill=black,opacity=0.980,line join=miter,line cap=butt,line width=0.800pt]
    (206.5178,46.9208) node[above right] (text23437) {$n$};
  \path[draw=black,line join=miter,line cap=butt,even odd rule,line width=0.800pt]
    (427.0892,34.8055) -- (423.1178,68.3988)(395.3450,49.3342) --
    (422.0308,68.6580) -- (471.4301,68.6582);
  \path[draw=black,fill=cffffff,line join=miter,line cap=butt,miter
    limit=4.00,nonzero rule,line width=0.800pt] (422.6590,69.0493) circle
    (0.4768cm);
  \path[draw=black,fill=cffffff,miter limit=4.00,line width=0.320pt]
    (419.2170,52.5460) -- (406.1570,65.6059)(405.8633,70.5131) --
    (424.1250,52.2515)(427.9469,53.0430) -- (406.6518,74.3381)(408.0607,77.5427)
    -- (431.1523,54.4511)(433.8095,56.4073) --
    (410.0165,80.2003)(412.3788,82.4514) -- (436.0618,58.7685)(437.8019,61.6418)
    -- (415.2529,84.1909)(418.6218,85.4354) --
    (439.0421,65.0151)(439.5468,69.1238) -- (422.7333,85.9373)(428.3718,84.9124)
    -- (438.5209,74.7632);
  \path[fill=black,line join=miter,line cap=butt,line width=0.800pt]
    (450.7623,96.3891) node[above right] (text23407) {$2$};
  \path[fill=black,line join=miter,line cap=butt,line width=0.800pt]
    (386.4556,46.9208) node[above right] (text23411) {$n$};
  \path[fill=black,line join=miter,line cap=butt,line width=0.800pt]
    (424.8309,30.2179) node[above right] (text23415) {$1$};
  \path[fill=black,line join=miter,line cap=butt,line width=0.800pt]
    (498.4774,62.0211) node[above right] (text23419) {$q_1$};
  \path[cm={{0.0,1.0,-1.0,0.0,(0.0,0.0)}},draw=black,fill=cffffff,miter
    limit=4.00,nonzero rule,line width=0.800pt]
    (75.4796,-495.7955)arc(-63.858:13.004:15.350826 and
    15.424)arc(13.004:89.866:15.350826 and 15.424)arc(89.866:166.728:15.350826 and
    15.424)arc(166.728:243.590:15.350826 and 15.424);
  \path[fill=black,line join=miter,line cap=butt,line width=0.800pt]
    (498.4774,80.1973) node[above right] (text23425) {$q_2$};
  \path[fill=black,line join=miter,line cap=butt,line width=0.800pt]
    (392.0676,93.2483) node[above right] (text23429) {$\ddots$};
  \path[draw=black,dash pattern=on 1.12pt off 0.56pt,line join=miter,line
    cap=butt,miter limit=4.00,even odd rule,line width=0.560pt] (489.7189,68.5220)
    -- (503.5181,68.5220);
  \path[draw=black,line join=miter,line cap=butt,even odd rule,line width=0.800pt]
    (453.3394,68.6122) -- (453.3394,85.7551);
  \path[fill=black,line join=miter,line cap=butt,line width=0.800pt]
    (435.5294,127.0625) node[above right] (text23571) {$1/z^{3}$};
  \path[fill=black,line join=miter,line cap=butt,line width=0.800pt]
    (249.7614,127.0625) node[above right] (text23575) {$1/z^{2}$};
  \path[draw=black,line join=miter,line cap=butt,even odd rule,line width=0.800pt]
    (67.3212,34.8055) -- (63.3498,68.3988) -- (45.3065,96.8114)(35.5770,49.3342)
    -- (62.2628,68.6580) -- (144.5754,68.6582);
  \path[draw=black,fill=cffffff,line join=miter,line cap=butt,miter
    limit=4.00,nonzero rule,line width=0.800pt] (62.8909,69.0493) circle
    (0.4768cm);
  \path[draw=black,fill=cffffff,miter limit=4.00,line width=0.320pt]
    (59.4490,52.5460) -- (46.3890,65.6059)(46.0953,70.5131) --
    (64.3569,52.2515)(68.1789,53.0430) -- (46.8838,74.3381)(48.2926,77.5427) --
    (71.3843,54.4511)(74.0415,56.4073) -- (50.2485,80.2003)(52.6108,82.4514) --
    (76.2937,58.7685)(78.0339,61.6418) -- (55.4849,84.1909)(58.8538,85.4354) --
    (79.2741,65.0151)(79.7788,69.1238) -- (62.9653,85.9373)(68.6037,84.9124) --
    (78.7529,74.7632);
  \path[fill=black,line join=miter,line cap=butt,line width=0.800pt]
    (146.5107,71.1178) node[above right] (text9357) {$2$};
  \path[fill=black,line join=miter,line cap=butt,line width=0.800pt]
    (61.7159,99.4411) node[above right] (text9361) {$\ldots$};
  \path[fill=black,line join=miter,line cap=butt,line width=0.800pt]
    (26.6876,46.9208) node[above right] (text9365) {$n$};
  \path[fill=black,line join=miter,line cap=butt,line width=0.800pt]
    (65.0629,30.2179) node[above right] (text9369) {$1$};
  \path[fill=black,line join=miter,line cap=butt,line width=0.800pt]
    (122.6755,50.5715) node[above right] (text9373) {$q_1$};
  \path[fill=black,line join=miter,line cap=butt,line width=0.800pt]
    (105.7433,50.5715) node[above right] (text9379) {$q_2$};
  \path[fill=black,line join=miter,line cap=butt,line width=0.800pt]
    (27.5228,77.0537) node[above right] (text9383) {$\vdots$};
  \path[draw=black,fill=cffffff,miter limit=4.00,nonzero rule,line width=0.800pt]
    (125.4764,54.0812)arc(-63.858:13.004:15.350826 and
    15.424)arc(13.004:89.866:15.350826 and 15.424)arc(89.866:166.728:15.350826 and
    15.424)arc(166.728:243.590:15.350826 and 15.424);
  \path[draw=black,dash pattern=on 1.12pt off 0.56pt,line join=miter,line
    cap=butt,miter limit=4.00,even odd rule,line width=0.560pt] (118.7689,57.1674)
    -- (118.7689,46.9730);
  \path[fill=black,line join=miter,line cap=butt,line width=0.800pt]
    (66.6509,127.0625) node[above right] (text23579) {$1/z$};

\end{tikzpicture}
\vspace{-7pt}
\caption{Wave-function renormalisation and tadpole graphs divergent in the forward limit.}
   \label{fig:spurious-div}
 \end{figure}

It is tempting to identify a single cut
$\Acut_{n+2} (\ell,p_{1},\dots,p_{n},-\ell)$
with the colour-stripped tree-level amplitude 
$\Atree_{n+2}(\ell,p_1,\dots,p_{n},-\ell)$ in the forward limit,
summed over the particle crossing the cut.
However, such a forward limit is divergent.
Indeed, if we choose to parametrise the $(n+2)$ on-shell momenta
with a complex parameter $z$ such that the last two momenta
\be
   q_2 = p_{n+1}(z)\xrightarrow[z \to 0]{}-\ell , \qquad
   q_1 = p_{n+2}(z)\xrightarrow[z \to 0]{} \ell ,
\label{forwardlimitparam}
\ee
depend linearly on $z$,
then there are two types of singular contributions:
\begin{enumerate}
\item single cuts of bubbles on external legs contain
      $1/(q_1+q_2+p_i)^2
      \xrightarrow[z \to 0]{} \mathcal{O}(1/z)$,
\item single cuts of massless tadpoles,
      depending on the propagator structure,
      such diagrams diverge as
      $1/(q_1+q_2)^2 \xrightarrow[z \to 0]{} \mathcal{O}(1/z^2)$,
      or even as
      $1/((q_1+q_2)^2(q_1+q_2+p_i)^2)
       \xrightarrow[z \to 0]{} \mathcal{O}(1/z^3)$.
\end{enumerate}
The corresponding diagrams are depicted in \fig{fig:spurious-div}
(also see e.g.~\cite[\S7.1]{Elvang:2013cua}).
Neither of these divergences can arise from the single cut,
because they correspond to cuts of the propagator renormalisation
or wave-function renormalisation of external legs. Such divergences cancel in supersymmetric gauge theories~\cite{Brandhuber:2005kd,CaronHuot:2010zt,ArkaniHamed:2010kv}.
Nevertheless, they are present in forward tree amplitudes
simply at the level of Feynman graphs.

One could in principle remove such diagrams~\cite{Ellis:2008ir,Ellis:2011cr},
at the expense of losing gauge invariance.
Another way is to regularise the forward limit~\cite{Britto:2011cr}
and then extract the regular part of $(n+2)$-point tree-level amplitude:
\begin{equation}
  \Atree_{n+2}(\ell, p_1,\dots,p_n,-\ell+z\hat q) =
  {\alpha_3 \over z^3}+{\alpha_2 \over z^2}+{\alpha_1 \over z} +
  \hAtree_{n+2}(\ell,p_1,\dots,p_n,-\ell)+\mathcal{O}(z) .  
\end{equation}
The precise value of $\hAtree_{n+2}$ depends on how the forward limit
is taken and is therefore scheme-dependent.
The forward-limit parametrisation that we use in the
massless case is detailed in \app{sec:forward-limit-param}.
However, it is clear that if we compute any unitarity cut
corresponding to a non-zero topology by cutting any further
propagators, then the divergent contributions and any scheme
dependence will be projected out.  In other words, such contributions
have zero physical cuts.  Therefore, we are allowed to identify the
regularised finite part of the forward tree amplitudes with the single
cut of the propagator $d_{0}=\ell^2+i\varepsilon=0$.  We thus have a
decomposition for the state-summed forward tree
\begin{equation}
   \sum_{h} (-1)^{2h} \hAtree_{n+2}(\ell^h,p_1,\dots,p_n,-\ell^{\bar{h}})
    = \sum_{r} \sum_{i_1<\dots<i_{r-1}}
      \frac{\Delta_{0,i_1,\dots,i_{r-1}}(\ell)}{d_{i_1}\!\dots d_{i_{r-1}}} .
\label{eq:forwardtreetoirrnums}
\end{equation}
Taking residues of the forward tree amplitude
with respect to the loci $d_{i_1}=d_{i_2}=d_{i_3}=0$
then gives the numerators $\Delta_{0,i_1,\dots,i_3}(\ell)$,
in the same way as the conventional quadruple cut~\cite{Britto:2004nc}
defined by $d_{0}=d_{i_1}=d_{i_2}=d_{i_3}=0$ of the one-loop amplitude.
Similarly, residues with respect to the loci $d_{i_1}=d_{i_2}=0$
match the triple cuts defined by $d_{0}=d_{i_1}=d_{i_2}=0$, and so on.
All the information
about the irreducible numerator of the one-loop amplitude is contained
in its (regularised) forward tree amplitudes.

The forward tree amplitude $\hAtree_{n+2}$ in \eqn{eq:forwardtreetoirrnums}
lacks contributions from the one-loop integral functions
that do not contain the propagator $d_0$ between $p_1$ and $p_n$.
There are hence only $\frac{(n-1)!}{r!(n-r-1)!}$ irreducible numerators
for $r$-gons in a given ordered forward amplitude.
Of course, the full set of one-loop numerators
(and the integral coefficients therein)
is recovered by considering all the external-leg permutations
of forward tree amplitudes.

\subsection{Single cut of one-loop string amplitudes}
\label{sec:cuts-string-ampl}

Cutkosky rules in superstring field theory have been recently
revisited in~\cite{Pius:2016jsl,Sen:2016uzq}, but not from a worldsheet
perspective. In this section we aim at doing so, as it will be needed
later in~\sec{sec:bcj-comp-monodr}. 

Intuitively, it is clear a single cut of a one-loop string theory
diagram should produce a diagram with tree-level topology and two
additional states. So we start by inserting a cutting operator in a
one-loop open-string amplitude of the form
$(\ell^2-m^2) \delta(\ell^2-m^2) \, \Theta(\ell^0)$. We will then
prove how this forces the integration of the moduli of the annulus to
localise onto the pinched annulus, which is a disk with two points
identified. In the ambitwistor
formulation~\cite{Mason:2013sva,Adamo:2013tsa,Ohmori:2015sha} of the
CHY formula~\cite{Cachazo:2013iea,Cachazo:2013hca}, this localisation
is automatically implemented by the loop-level scattering
equations~\cite{Geyer:2015bja,Geyer:2015jch,Geyer:2016wjx}. For us
here, the process is different --- we really look specifically at the
cut of a string amplitude.
 
We start with a generic bosonic open-string amplitude in the
representation where the loop momentum is not
integrated~\cite{DHoker:1988pdl} (the notation and conventions
follow~\cite{Tourkine:2016bak}) 
\beal\label{e:cAdef}
\cA(\alpha|\beta)= \int_0^\infty\!\!\!dt \int\!\dd^{D}\ell
\int_{\Delta_{\alpha|\beta}} \hspace{-15pt} d^{n-1}\nu \, &
e^{-\pi \alpha't \ell^2-2i \pi \alpha'\ell\cdot\sum_{i=1}^n p_i\,\nu_i} \\
\times \prod_{1\leq r<s\leq n}\!\!\! & f(e^{-2\pi t},\nu_r-\nu_s)
\times e^{-\alpha' p_r\cdot p_s\, G(\nu_r-\nu_s)} , 
\eeal
where the domain of integration $\Delta_{\alpha|\beta}$ is the union
of the domains of integration specified by 
$0\leq\Imm(\nu_{\alpha(1)})\leq \cdots\leq \Imm(\nu_{\alpha(p)})= t$ for
$\Ree(\nu_i)=0$  
and $t\geq \Imm(\nu_{\beta(p+1)})\geq \cdots \geq
\Imm(\nu_{\beta(n)})\geq0$ for $\Ree(\nu_i)=\frac12$. 
The non-zero mode part of the Green's function is given by
\begin{equation}\label{eq:G-def}
   G(\nu)=    -\log {\vartheta_1 (\nu|it)\over \vartheta_1'(0)}
=- \log {\sin(\pi \nu)\over
  \pi}-4\sum_{n\geq1} {q^n \over 1-q^n} {\sin^2(n\pi \nu)\over n} ,
\end{equation}
where we  have set $q=\exp(-2\pi t)$.
Note that since we have introduced the loop momentum,
there is no zero-mode contribution to this Green's function.

Now we consider the insertion of the single-cut operator
$\alpha'\ell^2\delta(\alpha'\ell^2)\,\Theta(\ell^0):=\alpha'\ell^2
\delta^{(+)}(\alpha'\ell^2)$ in the integrand,
and we shall show that the  integration localises at the pinched surface,
which is given by the infinitely long annulus for which $t=\infty$.
In the one-loop amplitude with the insertion of the single-cut operator,
\beal\label{e:cAproj}
   \cA(\alpha|\beta)^{\mathrm{proj}}:=
   \int\!\dd^{D}\ell\, \int_0^\infty\!\!\!dt \int_{\Delta_{\alpha|\beta}}  \hspace{-15pt} d^{n-1}\nu \, &
   e^{-\pi \alpha't \ell^2-2i \pi \alpha'\ell\cdot\sum_{i=1}^n p_i\nu_i}
   \times\alpha'\ell^2\delta^{(+)}(\alpha'\ell^2) \\
   \times \prod_{1\leq r<s\leq n}\!\!\! &
   f(e^{-2\pi t},\nu_r-\nu_s) \times e^{-\alpha' p_r\cdot p_s\,
   G(\nu_r-\nu_s)} ,
\eeal
the second line has a $q$-expansion
due to the excited string modes propagating in the loop.
Formally, integrating over the $\nu_i$ variables leads to an expression of the form
\begin{equation}
   \cA(\alpha|\beta)^{\mathrm{proj}} =
   \int\!\dd^{D}\ell\, \int_0^\infty\!\!\!dt\,  \alpha'\ell^2 \delta^{(+)}(\alpha'\ell^2)  \sum_{n\geq0}
c_n(t)\, e^{-\pi\alpha' \ell^2 t-2\pi n t} .
\end{equation}
We can then reabsorb the loop-momentum dependence of the exponential
by rescaling the proper time $t$:
\begin{equation}
   \cA(\alpha|\beta)^{\mathrm{proj}} =
   \int_0^\infty\!\!\!dt \int\!\dd^{D}\ell\,
   \sum_{n\geq0}\frac{\alpha'\ell^2 \delta^{(+)}(\alpha'\ell^2)}
                     {\alpha'\ell^2-2n}
   c_n\left(\frac{t}{\alpha'\ell^2-2n}\right)\, e^{-\pi  t} .
\end{equation}
The delta-function insertion projects the integral on the $n=0$ sector,
\begin{equation}
   \cA(\alpha|\beta)^{\mathrm{proj}} =
   \int_0^\infty\!\!\!dt \int\!\dd^{D}\ell\,\delta^{(+)}(\alpha'\ell^2)\,
   c_0\left(\frac{\alpha' t}{\alpha'\ell^2}\right)\, e^{-\pi  t} .
\end{equation}

This final expression is identical to what would have happened if we
had chosen the leading term $\mathcal{O}(1)$ term in the $q$-expansion. This is
effectively equivalent to having set $q=0\Leftrightarrow t=+\infty$ in
the one-loop single-cut string integrand, which eventually proves our
initial claim.

\section{Momentum kernel in the forward limit}
\label{sec:monodromy}
In this section we study the monodromy relations satisfied
by the colour-stripped forward tree-level amplitudes in gauge theory.

\subsection{Momentum kernel}

Colour-stripped open-string disc amplitudes
(denoted by a calligraphic $\cA$) satisfy the following
fundamental monodromy relation~\cite{Plahte:1970wy}
(with $p_{r\cdots s}=\sum_{i=r}^s p_i$)
\be
\label{e:Mon1}
   \cAtree_{n+2}(p_1,p_2,\dots,  p_{n+2})
    + \sum_{i=2}^{n+1} e^{i\alpha^\prime p_1\cdot p_{2\cdots i}}
      \cAtree_{n+2}(p_2,\dots,\!\!\!
      \underbrace{p_1}_{\textrm{position}~i}\!\!\!,\dots,p_{n+2}) = 0 .
\ee
Such relations, obtained by circulating a single momentum $p_i$,
generate all the monodromy relations between the open string
amplitudes~\cite{BjerrumBohr:2009rd,BjerrumBohr:2010zs,Stieberger:2009hq}. 
The external states can be massive or massless, because,
as explained in~\cite{Bjerrum-Bohr:2013bxa},
monodromy relations between tree amplitudes
are generic properties common to any tree amplitude
independently of the details of the theory.

From now we set $q_1=p_{n+2}$ and $q_2=p_{n+1}$.  These momenta can be
massless or massive and  satisfy the momentum conservation relation 
\begin{equation}
   p_1+\cdots+p_n= -q_1-q_2 .
\end{equation}

The amplitudes being real, taking the real and imaginary part of
(\ref{e:Mon1}) simply amounts to taking cosines or sines from the
phases.
In the limit of infinite string tension,  $\alpha'\to0$, one obtains
relations between colour-ordered field theory amplitude
$A(q_1,p_1,\dots,p_n,q_2)$ and its external-leg permutations.

The leading order-$\alpha'$ contribution of the real part of~\eqref{e:Mon1},
using that $\cos(\alpha'p\cdot q)\simeq1$,
leads to the photon-decoupling identities 
\begin{subequations}
\begin{align}
   \Atree_{n+2}(q_1,p_1,p_2,\dots,p_n,q_2)
    + \sum_{i=2}^{n+1} \Atree_{n+2}(p_2,\dots,\!\!\!
      \underbrace{p_1}_{\textrm{position}~i}\!\!\!,\dots,p_n,q_2,q_1)=0 ,
\label{e:photon1} \\
   \Atree_{n+2}(q_1,p_1,p_2,\dots,p_n,q_2)
    + \sum_{i=2}^{n+1} \Atree_{n+2}(p_1,\dots,\!\!\!
      \underbrace{q_1}_{\textrm{position}~i}\!\!\!,\dots,p_n,q_2)=0 .
\label{e:photon2}
\end{align}
\label{e:photon}%
\end{subequations}
Furthermore, since $\sin(\alpha' p\cdot q)\simeq\alpha' p\cdot q$,
the imaginary part of~\eqref{e:Mon1} at leading order in $\alpha'$
leads to so-called fundamental BCJ relations~\cite{Feng:2010my}
that imply the full range of the BCJ kinematic relations~\cite{Bern:2008qj} between gauge-theory amplitudes.
Circulating the momentum $p_1$ 
\begin{equation}\label{e:FundBCJ1}
  \sum_{i=1}^n\, p_1\cdot (q_1+p_{2\cdots i})\,
  \Atree_{n+2}(p_2,\dots,\!\!\!
  \underbrace{p_1}_{\textrm{position}~i}\!\!\!,\dots,p_n,q_2,q_1)=0 ,
\end{equation}
or circulating $q_1$
\begin{equation}\label{e:FundBCJ2}
  \sum_{i=1}^n\, q_1\cdot p_{1\cdots i}\,
  \Atree_{n+2}(p_1,\dots,\!\!\!\!\!\!
  \underbrace{q_1}_{\textrm{position}~i+1}\!\!\!\!\!\!,\dots,p_n,q_2)=0 ,
\end{equation}
and circulating $q_2$
\begin{equation}\label{e:FundBCJ3}
 q_2\cdot q_1\, \Atree_{n+2}(q_1,q_2,p_1,\dots,p_n)+
  \sum_{i=1}^{n-1}\, q_2\cdot (q_1+p_{1\cdots i})\,
  \Atree_{n+2}(p_1,\dots,\!\!\!\!\!\!
  \underbrace{q_2}_{\textrm{position}~i+1}\!\!\!\!\!\!,\dots,p_n,q_1)=0 .
\end{equation}
These relations generate all the monodromy BCJ relations satisfied by
the colour-ordered amplitudes.
The resulting equations are concisely rewritten using the momentum-kernel
formalism~\cite{BjerrumBohr:2010hn}.
The power of this formalism is that in addition to the BCJ relations
it provides simultaneous and gauge-invariant
treatment of the KLT construction~\cite{Kawai:1985xq}
for gravity amplitudes out of gauge-theory amplitudes.
Because of the two marked momenta $q_1$ and $q_2$ we have three kinds
of momentum kernels and corresponding kinematic relations.
\begin{itemize}
\item
Monodromy relations acting only on an external momentum $p_i$
and neither $q_1$ nor $q_2$:
\begin{equation}\label{e:SAtree}
   \sum_{\sigma\in\mS_{n}} \cS
   [\sigma(1,\dots,n)|\beta(1,\dots,n)]_{q_2}
   \Atree_{n+2}(q_1,\sigma(p_1,\dots,p_n),q_2) = 0 ,
\end{equation}
where $\mS_n$ is the set of permutations of $n$ elements,
$\beta$ is any permutation of legs $1,\dots,n$,
and the sum runs over all permutations $\sigma$ of these legs.
The momentum kernel $\cS$ is given
by~\cite{Bern:1998sv,BjerrumBohr:2010ta,BjerrumBohr:2010hn}\footnote{
  There is some freedom in the expression for the momentum
  kernel due to the various different ways of organising the KLT
  relation between closed-string amplitudes and open-string
  amplitudes~\cite{Kawai:1985xq}. In this work we follow the contour
  deformation used in~\cite{BjerrumBohr:2010hn} which leads to flip in
  the ordering of the legs in the right-moving amplitude compared to the
  left moving amplitude as given in~\cite[eq.~(2.16)]{BjerrumBohr:2010hn}.
  A different
  ordering of the leg in the right-moving amplitude will result in a
  different form for the momentum kernel as used for example
  in~\cite{Broedel:2013tta,Cachazo:2013iea,Mafra:2016ltu}. These
  different forms are equivalent since they lead to the equivalent
  linear relations between the colour-ordered gauge theory amplitudes,
  and the same gravitational
  amplitudes.}
\begin{equation}\label{e:Sdef}
   \cS[\sigma(1,\ldots,n)|\beta(1,\ldots,n)]_q := \prod_{i=1}^n\,
      \Big(q\cdot p_{\sigma(i)} + \Theta_i(\sigma,\beta)\Big) ,
\end{equation}
where the quantity $ \Theta_i(\sigma,\beta)$ is defined by
\begin{equation}\label{e:ThetaiDef}
  \Theta_i(\sigma,\beta)= \sum_{j=i+1}^n \theta\big(\sigma(i),\beta(j)\big)\,
  p_{\sigma(i)}\cdot p_{\beta(j)} .
\end{equation}
Here $\theta(i,j)$ is 1 if the ordering of the legs $i$ and $j$ is
opposite in the sets $\{i_1,\dots,i_k\}$ and $\{j_1,\dots,j_k\}$ and 0
if the ordering is the same. Moreover, the reference momentum~$q$
should not belong to the set $\{1,\dots,n\}$, hence
$\Theta_i(\sigma,\beta)$ does not depend on $q_1$.
\item The monodromy relations acting on one of the special momentum,
say $q_1$,
\begin{equation}\label{e:SAtreebis}
\sum_{\sigma\in\mS_{n}} \cS
[\sigma(q_1,2,\dots,n)|\beta(q_1,2,\dots,n)]_{q_2}
\Atree_{n+2}(p_1,\sigma(q_1,p_2,\dots,p_n),q_2) = 0 .
\end{equation}
\item The monodromy relations moving both special momenta $q_1$ and $q_2$
\begin{equation}\label{e:SAtreeter}
\sum_{\sigma\in\mS_{n}} \cS
[\sigma(q_2,q_1,2,\dots,n-1)|\beta(q_2,q_1,2,\dots,n-1)]_{p_1}
\Atree_{n+2}(p_n,\sigma(q_2,q_1,p_2,\dots,p_n),p_1) = 0 .
\end{equation}
\end{itemize}

\subsection{Forward limit of the fundamental monodromy relation}
\label{sec:forw-limit-fund}

Now we consider the forward limit of the monodromy relation~\eqref{e:SAtree}.
It is obtained by taking the limit $q_1+q_2\to 0$ so
that $p_1+\cdots+p_n\to0$. We set $q_1\to\ell$ and $q_2\to-\ell$.
Care must be exercised when taking the forward limit as the tree
amplitudes can develop divergences. 

If the forward limit is parametrised by a parameter $z\to0$,
the discussion of \sec{sec:part-integr-single}
implies that the tree amplitude develops at most third-order poles in $z$:
\begin{equation}
  \lim_{z\to 0} \Atree_{n+2}(q_1, p_1,\dots,p_n,q_2)=
  {\alpha_3\over z^3}+ {\alpha_2\over z^2}+{\alpha_1\over z} +
  \hAtree_{n+2}(\ell,p_1,\dots,p_n,-\ell)+\mathcal O(z) .  
\end{equation}
The momentum kernel is a polynomial in the loop momentum.
It does not have poles and has a $z$-expansion of at most second order in $z$ for fundamental monodromy relations,
and at most of the order $z^{2n}$ for the momentum kernel in \eqn{e:Sdef}. 
The forward limit of the monodromy relation~\eqref{e:SAtree} reads
\begin{equation}
  \left( \mathcal S(\ell)+ z \mathcal S^{(1)}(\ell)+z^2
    \mathcal S^{(2)}(\ell)+z^3
    \mathcal S^{(3)}(\ell)+\mathcal O(z^4)\right) \, \left( {\alpha_3\over z^3}+ {\alpha_2 \over
      z^2}+{\alpha_1\over z}+\hAtree_{n+2}+\mathcal O(z)\right) = 0  ,
\end{equation}
leading to the system of equations
\begin{subequations}
\begin{align}
  \mathcal S(\ell)\,\alpha_3 &= 0 , \\
  \mathcal S(\ell)\,\alpha_2+\mathcal S^{(1)}(\ell)\, \alpha_3 &= 0 ,\\
  \mathcal S(\ell)\,\alpha_1+\mathcal S^{(1)}(\ell)\, \alpha_2+\mathcal S^{(2)}(\ell)\,\alpha_3 &=0 , \\
  \mathcal S(\ell)\, \hAtree+\mathcal S^{(1)}(\ell)\, \alpha_1+\mathcal S^{(2)}(\ell)\,\alpha_2+\mathcal S^{(3)}(\ell)\,\alpha_3 &= 0.
\end{align}
\end{subequations}

The last equation shows that the finite part of forward tree amplitude
satisfies an inhomogeneous monodromy relation of the form
\begin{equation}
   \mathcal S(\ell)\, \hAtree_{n+2} = R_{n+2}(\ell) .
\end{equation}
where $R_{n+2}$ is defined in terms of the $\mathcal{S}^{(i)}$ and $\alpha_{i}$.
For the case of the fundamental BCJ monodromy relation we have
\begin{equation}\label{e:FundBCJReg}
  \sum_{i=1}^n\, p_1\cdot (\ell+p_{2\cdots i})\,
  \hAtree_{n+2}(p_2,\dots,\!\!\!
  \underbrace{p_1}_{\textrm{position}~i}\!\!\!,\dots,p_n,-\ell,\ell)=R_{n+2}(p_1,\dots,p_n,\ell) ,
\end{equation}
with equivalent statements, mutatis mutandis, for the forward limits of~\eqref{e:SAtreebis} and~\eqref{e:SAtreeter}.

Since the modified relation~\eqref{e:FundBCJReg} depends on the
regularisation used to extract the finite part, we need to analyse the
effect of the regularisation on the monodromy relations.
As explained in \sec{sec:part-integr-single}, the
residues of the regularised forward tree amplitudes give the same
residues from the multiple cut of the one-loop amplitude.
The right-hand side of the regularised monodromy relation~\eqref{e:FundBCJReg} does not
contribute the residues because the divergences arise only
from the tadpole and wave-function renormalisation contributions
in \fig{fig:spurious-div}.  
Therefore the monodromy relations between the various residues
and thus the irreducible numerators are independent of
the regularisation and non-ambiguous.

It is interesting to note that there exists a regularisation for which $R_{n+2}(p_1,\dots,p_n,\ell)=0$.
This is the one used for the partial amplitudes $a(p_1,\dots,p_n,-,+)$
in~\cite{Geyer:2015jch,He:2016mzd,He:2017spx}  which are regulated using the $\mathcal
Q$-cut prescription~\cite{Baadsgaard:2015twa} or the CHY
prescription~\cite{Geyer:2015jch,He:2015yua,Cachazo:2015aol}. It is shown 
in these works that the partial amplitudes satisfy the fundamental BCJ
monodromy relations~(\ref{e:FundBCJ1}) with a vanishing  right-hand side.
Moreover, $\mathcal N=4$ super-Yang-Mills
amplitudes are completely free of forward-limit singularities~\cite{Brandhuber:2005kd,CaronHuot:2010zt,ArkaniHamed:2010kv}.
From now on we will assume that our forward tree amplitudes are
regularised in the way that preserves the form of the fundamental BCJ relations.

\subsection{Fundamental monodromy relations at one loop}
\label{sec:fund-BCJ-oneloop}

Now we assume that the fundamental monodromy relations
in~\eqn{e:FundBCJ1} (and its permutations) hold for the forward tree amplitudes:
\begin{subequations}
\begin{align}
\label{e:FundBCJF1}
   \sum_{i=1}^n\, p_1\cdot (\ell+p_{2\cdots i})\,
   \hAtree_{n+2}(p_2,\dots,
      \!\!\!\underbrace{p_1}_{\textrm{position}~i}\!\!\!,
      \dots,p_n,-\ell,\ell)&=0 ,\\
\label{e:FundBCJF2}
   \sum_{i=1}^n\, \ell\cdot p_{1\cdots i}\,
   \hAtree_{n+2}(p_1,\dots,
      \!\!\!\!\!\!\underbrace{\ell}_{\textrm{position}~i+1}\!\!\!\!\!\!,
      \dots,p_n,-\ell)&=0 ,\\
\label{e:FundBCJF3}
   \sum_{i=1}^{n-1}\,\ell \cdot (\ell+p_{1\cdots i})\,
   \hAtree_{n+2}(p_1,\dots,
      \!\!\!\!\!\!\underbrace{-\ell}_{\textrm{position}~i+1}\!\!\!\!\!\!,
      \dots,p_n,\ell)&=-\ell^2\, \hAtree_{n+2}(\ell,-\ell,p_1,\dots, p_n) .
\end{align}
\end{subequations}

Since at one loop there are only two special legs, with the incoming
and outgoing loop momentum, one can find monodromy relations amongst
single cuts of planar graphs only.  The forward limits of the
monodromy relations in \eqns{e:FundBCJF2}{e:FundBCJF3} mix single cuts
of planar and non-planar amplitudes.  But since one can always express
the field-theory non-planar amplitudes as a combination of planar
amplitudes~\cite{Bern:1990ux}, we do not have to study these equations
in too much detail.  Of course, at higher loops the non-planar
contributions cannot be avoided~\cite{Tourkine:2016bak}.\footnote{The
  integrand monodromy relations in string
  theory~\cite{Tourkine:2016bak,Hohenegger:2017kqy} relate planar and
  non-planar amplitudes by distributing the external legs on the
  various boundaries of the open-string amplitude.  It should be
  noted that these relations do not change the master topology of
  the open-string vacuum graphs but just the position of the external
  legs inside a given topology.}

An explicit solution of the monodromy relations between tree
amplitudes expressing the amplitudes
$\hAtree_{n+2}(\ell,\alpha,p_1,\beta,-\ell)$
in the minimal basis of $(n-3)!$ amplitudes
with leg 1 next to $\ell$,
$\hAtree_{n+2}(\ell,p_1,\gamma,-\ell)$,
can be read off from \cite[eq.~(4.22)]{Bern:2008qj}:
\begin{equation}
   \hAtree_{n+2} (\ell,\alpha,p_1,\beta,-\ell) = \!\!
      \sum_{\sigma \in S(\alpha) \shuffle \beta} \!\!\!
      \hAtree_{n+2} (\ell,p_1,\sigma,-\ell) \prod_{i=1}^{|\alpha|}
      \frac{ {\cal F}_\ell(\sigma,\alpha_i) }
           { s_{\ell \alpha_1 \dots \alpha_i} } ,
\label{e:bcjAcorner}
\end{equation}
where $\alpha$ is a short-hand notation for the permutations of the
external legs $p_{\alpha(2)},\dots,p_{\alpha(r)}$, $\beta$~for the permutations of the
external legs $p_{\beta(r+1)},\dots,p_{\beta(n)}$. The sum is over 
the permutations~$\gamma$ running over the shuffle products between $\beta$
and the permutations of~$\alpha$.  There are at most $(n-1)!$
terms in the expansion in agreement with the dimension of the minimal
basis for colour-ordered amplitudes.
The kinematic factor ${\mathcal F}_\ell(\sigma,\alpha(i)) $ is given by
\begin{equation}\!\!\!\!
{\mathcal F}_\ell(\sigma,\alpha_i) =
    - s_{1\alpha_i}
    - \!\!\!\!\!\sum_{ \sigma_j \in\;\!\overline{\sigma}^i \cap
                       (\underline{\alpha}_i \cup \beta) }\!\!\!\!\!
      s_{\alpha_i \sigma_j}
    - \left\{\!
\begin{array}{ll}
    - s_{\ell\alpha_1\dots\alpha_{i-1}} &
      \mbox{if $\sigma^{-1}_{\alpha_{i-1}} < \sigma^{-1}_{\alpha_i} < \sigma^{-1}_{\alpha_{i+1}}$} \\
      s_{\ell\alpha_1\dots\alpha_i} & 
      \mbox{if $\sigma^{-1}_{\alpha_{i-1}} > \sigma^{-1}_{\alpha_i} > \sigma^{-1}_{\alpha_{i+1}}$} \\
      s_{\ell\alpha_i}
    + \sum_{\alpha_j \in \overline{\alpha}^i} s_{\alpha_i \alpha_j} &
      \mbox{if $\sigma^{-1}_{\alpha_i} > \sigma^{-1}_{\alpha_{i-1}} , \sigma^{-1}_{\alpha_{i+1}}$} \\
      0 & \mbox{else}
\end{array}\!\right\} .\!
\label{e:Ffunctionl}
\end{equation}
Here by $\sigma^{-1}_{\alpha(i)}$ we denote the position of leg $\alpha(i)$ in the set $\sigma$,
and $\overline{\sigma}^i$ and $\underline{\sigma}_i$ are the subsets of $\sigma$
comprising the elements that precede or follow $\alpha(i)$, respectively.
The special cases  for  non-existent elements of $\alpha$
as follows
\begin{equation}
   \sigma^{-1}_{\alpha(0)} := \sigma^{-1}_{\alpha(2)} , \qquad \quad
   \sigma^{-1}_{\alpha({|\alpha|+1})} := 0 .
\end{equation}

Here we have adapted the expression given in~\cite{Bern:2008qj}
to render the dependence on legs $\ell$ and $-\ell$ explicit.
Other colour-ordered amplitudes are mapped to
the minimal basis by a combination of Kleiss-Kuijf relations~\cite{Kleiss:1988ne,DelDuca:1999rs}
and the monodromy relation given above.
Note that although
this solution's denominators are not divergent in the forward limit,
this is not true for the entirety of the momentum kernel.

\subsection{Forward limit of the momentum kernel }
\label{sec:moment-kern-forw}

In the forward limit where $q_1=-q_2=\ell$ so that $p_1+\cdots+p_n=0$,
the momentum-kernel relation~\eqref{e:Sdef} is a polynomial of
degree $n$ in $\ell$ of degree at most linear in each $\ell\cdot p_i$
with $1\leq i\leq n$,
\begin{equation}\label{e:SS}
   \cS(\ell)_{\sigma,\beta}:=
   \cS[\sigma(1,\dots,n)|\beta(1,\dots,n)]_{\ell} =
   \prod_{i=1}^n \left(\ell\cdot p_{\sigma(i)}+\Theta_i(\sigma,\beta)
                 \right) ,
\end{equation}
where $  \Theta_i(\sigma,\beta)$ is defined in~\eqref{e:ThetaiDef},
and it is important that this quantity 
does not depend on the momentum $\ell$.  
Since in the forward limit the reference momentum is $\ell$ and
permutations are acting on all the $n$ other external legs,
we use a short-hand matrix notation $\cS(\ell)_{\sigma,\beta}$.

In the forward limit, we find that
due to the kinematical constraint $p_1+\cdots+p_n=0$
the momentum kernel degenerates, and its rank decreases.
For massless momenta $p_i^2=0$
we computed numerically that 
\begin{equation}\label{e:dimKerF}
   \textrm{dim}~\text{Ker}(\cS(\ell))
     = (n-3)! \, (n^2-2n+2)=(n-1)!+(n-2)!+2(n-3)!,
\end{equation}
see \tab{tab:Skernel-dim}.
Moreover, we find the same dimension for the forward limit of the full momentum kernel in string theory
\begin{equation}
      \cS_{\alpha'}(\ell)= \prod_{i=1}^n \sin\left(\alpha'\left(\ell\cdot p_i+\Theta_i(\sigma,\beta)\right)\right) .
\end{equation}

\begin{table}[t]
   \centering
\begin{tabular}{|l|llllll|l|}
\hline
   \text{n} & $4$ & $5$ & $6$ & $7$ & 8 & 9 &  generic $n$ \\
\hline
   \text{dim Ker} $\cS_{n+2}|_{q}$
          & 6 & 24 & 120 & 720 & 5040 & 40320& $(n-1)!$ \\
   \text{dim Ker} $\cS(\ell)$~\eqref{e:SS}
          & 10 & 34 & 156 & 888 & 6000 & 46800 &$(n-3)!(n^2-2n+2)$ \\
   \text{dim Ker} $\cS$~\eqref{e:SS2}~\textrm{or}~\eqref{e:SS3}
          & 2 & 4 & 12 & 48 & --- & ---& $2(n-3)!$ \\
\hline
\end{tabular}
\caption{Dimension of the kernel of the momentum kernel at $n+2$
points and in the forward limit. The first line gives the dimension of
the kernel before taking the forward limit. We have numerically checked the rank for $n=4,5,6$ and $7$. Since $\textrm{dim}~\text{Ker}~\eqref{e:SS}/(n-3)!$ cannot be more than a quadratic polynomial,
three numerical points are enough to confirm the formula.}
\label{tab:Skernel-dim}
\end{table}

We propose the following interpretation.  There are $(n+2)!$
different orderings of the amplitudes including the permutations of
the special forward legs $\ell$ and $-\ell$.  In the forward limit
some monodromy transformations vanish and leave more independent
amplitudes characterised by the position of these special legs.
Before taking the forward limit one expresses all colour-ordered
amplitudes as a linear combination of elements in the basis composed
of $\Atree_{n+2}(q_1,p_1,\alpha,q_2)$, where
$\alpha\in\mathfrak S_{n-1}$ is the set of permutations of $n-1$
elements.  In the forward limit this basis is composed of the single
cuts of planar one-loop graphs
\begin{equation}
\label{e:Basis1} 
   B_{n-1}(\alpha) :=  \hAtree_{n+2}(\ell,p_1,\alpha,-\ell),
   \qquad\qquad~~\alpha\in\mathfrak S_{n-1} .
\end{equation}

Normally, for any choice of three momenta $k_1, k_2$ and $k_3$
amongst $\{q_1,q_2,p_1,\dots,p_n\}$,
any colour-ordered tree amplitude
$\Atree_{n+2}(k_1,\alpha,k_2,\beta,k_3,\gamma)$,
where the disjoint union of the permutations $\alpha$, $\beta$ and $\gamma$ runs over the the permutation of $n-1$ momenta $\mathfrak S_{n-1}$,
can be expanded
in the minimal basis $\Atree_{n+2}(k_1,k_2,\sigma,k_3)$, using a
combination of Kleiss-Kuijf
relations~\cite{Kleiss:1988ne,DelDuca:1999rs} and the BCJ mapping
given in~\cite[eq.~(4.22)]{Bern:2008qj}. 

In the forward limit one needs to make the forward momenta special,
and the choice of the three fixed momentum affects the monodromy relations.
Some forward tree amplitudes become independent,
depending whether the forward loop momenta are fixed or not.
This is due to the vanishing of some coefficients in the monodromy
relations. This happens when the BCJ
map in~\cite[eq.~(4.22)]{Bern:2008qj} develops a pole.
The following two classes of regularised forward tree amplitudes
cannot be related by a BCJ transformation to the element of basis~\eqref{e:Basis1}
\begin{subequations}
\begin{align}
\label{e:Basis2} &
B_{n-2}(\alpha) :=\hAtree_{n+2}(\ell,p_1,\alpha,-\ell,p_n),
   \qquad\quad\:\,\alpha\in \mathfrak S_{n-2},\\
\label{e:Basis3} &\!
\begin{aligned}
B^1_{n-3}(\alpha) :=\hAtree_{n+2}(p_1,p_2,\alpha,\ell,-\ell,p_n),
   \qquad\alpha\in\mathfrak S_{n-3}, \\
B^2_{n-3}(\alpha) :=\hAtree_{n+2}(p_1,p_2,\alpha,-\ell,\ell,p_n),
   \qquad\alpha\in\mathfrak S_{n-3},
\end{aligned}
\end{align}
\end{subequations}
There are $(n-1)!$ elements in~\eqn{e:Basis1}, $(n-2)!$ elements
in~\eqn{e:Basis2} and  $2(n-3)!$ elements in \eqn{e:Basis3}.  
The sum of these  dimensions  equals to the one  given
in~\eqn{e:dimKerF}. Notice that all the  forward amplitudes in the sets
$B_{n-1}(\alpha)$, $B^1_{n-3}(\alpha,)$ and
$B^2_{n-3}(\alpha)$
arise from the single cut of planar one-loop amplitudes.  The
amplitudes in the set $B_{n-2}(\alpha)$ arise from the single cut of
non-planar one-loop amplitudes.

The dimension in~\eqn{e:dimKerF} provides an upper bound on the number of partial one-loop amplitudes. 
For purely gluonic amplitudes in QCD and $\mathcal N=4$
super-Yang-Mills amplitudes at one loop, the planar cuts are enough.
We can always consider single cuts that can be
expressed on the minimal basis~\eqref{e:Basis1}.
For amplitudes with multiple flavoured matter particles
in the fundamental representation,
it is not yet entirely clear if non-planar cuts may be avoided as well.
It would be interesting to  determine the interplay with
colour-kinematics duality for fundamental
matter~\cite{Johansson:2014zca,Johansson:2015oia} and understand
if some single cuts need to be expressed in the
basis~(\ref{e:Basis2})--(\ref{e:Basis3}).  For instance, the Kleiss-Kuijf relations~\cite{Kleiss:1988ne,DelDuca:1999rs,Johansson:2014zca,Johansson:2015oia}
on non-planar cuts involve the planar cuts in~(\ref{e:Basis3}). 
We note, as well, that the monodromy relations at higher-loop mix planar and  non-planar
cuts.

For the amplitude relation~\eqref{e:SAtreebis} the momentum kernel is 
\begin{equation}\label{e:SS2}
\cS[-\ell,2,\dots,n|\beta(-\ell,2,\dots,n)]_{\ell}=
\Big(-\ell^2-\sum_{i=2}^n\ell\cdot p_i\, \theta(1,i)\Big)\,\prod_{i=2}^n
\Big(\ell\cdot p_i+\sum_{j>i}^n p_i\cdot p_j\, \theta(i,j)\Big)
\end{equation}
and for the relations~\eqref{e:SAtreeter} the momentum kernel is 
\begin{multline}\label{e:SS3}
\cS[\ell,-\ell,3,\dots,n|\beta(\ell,-\ell,3,\dots,n)]_{p_1}=\prod_{i=3}^n
\Big(p_1\cdot p_i+\sum_{j>i}^n p_i\cdot p_j \,\theta(i,j)\Big) \cr
\times
\Big(p_1\cdot \ell-\ell\cdot\ell\, \theta(1,2)+ \sum_{j=3}^n \ell\cdot p_j
\theta(1,j)\Big) \Big(-p_1\cdot\ell-\sum_{j=3}^n \ell\cdot p_j \,\theta(2,j) \Big) .
\end{multline}
We recall that $\theta(i,j)$ is defined below~\eqn{e:ThetaiDef}.
We find that both these momentum kernel have dimension $2(n-3)!$.
This is  the dimension of the basis of amplitudes in $B^1_{n-3}(\alpha)$ and
$B^2_{n-3}(\alpha)$ in~\eqref{e:Basis3}.

\subsection{Monodromy relations and one-loop coefficients}
\label{sec:one-loop-amplitude}

The monodromy relations imply that we can express the forward tree
amplitudes in the minimal basis of forward tree amplitudes $\mathcal B=\{B^{\rm tree}_r(\ell)\}$
\begin{equation}
   \hAtree_{n+2} (\sigma (\ell,p_{1},\dots,p_{n},-\ell)) =
   \sum_{r=1}^{\textrm{dim}(\mathcal B)}
   c^\sigma_r(\ell) \, B^{\rm tree}_r(\ell) .
\end{equation}
The coefficients $c_r^\sigma(\ell)$ are explicitly known by solving
the solving the monodromy relations (cf.~\cite[eq.~(4.22)]{Bern:2008qj}).

The dimension of the kernel of the momentum kernel $\mathcal S(\ell)$
in~\eqref{e:SS} gives an upper bound on the dimension of the minimal basis 
 dim$(\mathcal B)\leq (n-3)! (n^2-2n+2)$.
The forward tree amplitudes $ \hAtree_{n+2} (\ell,\sigma(p_{1},\dots,p_{n}),-\ell)$
can be expressed in the basis~\eqref{e:Basis1} of dimension
$(n-1)!$. This is enough for the single cuts of planar
one-loop amplitudes.

We have explained in \sec{sec:part-integr-single} that the
forward tree amplitudes contain all the information about the
irreducible one-loop numerators $\Delta_{i_1,\dots,i_r}(\ell)$. 
Therefore the number of independent one-loop irreducible numerators is
given by the number of independent numerators in the minimal basis of forward tree
amplitudes. 
Each of the numerators contains a one-loop integral coefficient
that is non-spurious. So far our analysis provides an  upper bound on
the number of one-loop integral coefficients, it would be interesting
to refine this analysis to obtain the optimal number of kinematically
independent coefficients along the lines of~\cite{Chester:2016ojq,Primo:2016omk}.

\section{Field-theory monodromy relations from string theory}
\label{sec:bcj-comp-monodr}

In this section, we turn to the
one-loop string-theory monodromies of~\cite{Tourkine:2016bak} and their
field-theory incarnation~\cite{Boels:2011tp,Boels:2011mn}, see also~\cite{Du:2012mt}.

In the previous sections we have studied the kinematic relations on
single-cuts obtained from the monodromies of forward tree-amplitudes.
In \sec{sec:loopvsforward} we re-derive the forward-tree monodromies 
starting from the string-theory loop-monodromies.  In
\sec{sec:dissecting}, we describe an important consequence of the
string theory monodromies concerning the colour-kinematic
duality. First we explain how they actually give exact information
about integrands even before integration. We do this as a first step
toward a better control over the ambiguities in the labelling of the
loop momentum when considering integration-by-parts
identities~\cite{Tkachov:1981wb,Chetyrkin:1981qh,Laporta:1996mq,Laporta:2001dd}.
We then illustrate this point by explaining how the BCJ relations are
compatible with these field-theory monodromy relations obtained from
string theory.

\subsection{Loop integrand monodromy versus forward-amplitude monodromies}
\label{sec:loopvsforward}

The field-theory limit of the integrand relations in string theory
leads to the relations
that were previously derived using field-theory techniques
in~\cite{Boels:2011tp,Boels:2011mn}.
The first-order string-theory contributions
are discussed in detail in \app{sec:one-loop-monodromies}.

The one-loop fundamental monodromy relations between planar and
non-planar open string integrands, $\mathcal A(p_{i_1},\dots,p_{i_r})$
and $\mathcal A(p_{i_1},\dots,p_{i_r}|p_{i_1},\dots,p_{i_s})$ reads
\begin{equation}\label{e:MonString}
  \sum_{i=1}^{n-1} e^{i\pi \alpha' p_1\cdot (\sum_{j=1}^i p_j)}
  \, \mathcal A(p_2,\dots,\!\!\!
  \underbrace{p_1}_{\textrm{position}~i}\!\!\!,\dots,p_{n})=
  e^{-i\pi\alpha'\ell\cdot p_1}\mathcal
  A(p_2,\ldots,p_n|p_1) .
\end{equation}
It was argued in~\cite{Tourkine:2016bak} at the first order in
$\alpha'$ these relationship lead to the one-loop planar integrand
relations in field theory
\begin{equation}\label{e:MonQFT}
\sum_{i=1}^{n-1} p_1\cdot (\ell+\sum_{j=1}^{i} p_j) \, \mathcal
I(p_2,\dots,\!\!\!
\underbrace{p_1}_{\textrm{position}~i},\!\!\!\dots,p_{n})\approx 0 .
\end{equation}
The notation ``\,$\approx 0$'' means that, in field theory, the
relations are valid up to contributions that vanish upon integration
of the loop momentum~\cite{Boels:2011tp,Boels:2011mn}.  We show in
\sec{sec:role-boundary-terms} that the string-theoretic construction
actually determines the form of the terms in the right-hand side of
\eqn{e:MonQFT}. This implies that the relations can be thought of as
exact integrand relationships.

At this stage, \eqn{e:MonQFT} seems different from the
forward tree monodromy relation~\eqref{e:FundBCJF1}.
The latter involves $n$ terms,
whereas the former contains only $n-1$.
So, first of all, we need to show how to relate both constructions.

\begin{figure}[t]
  \centering
  \input{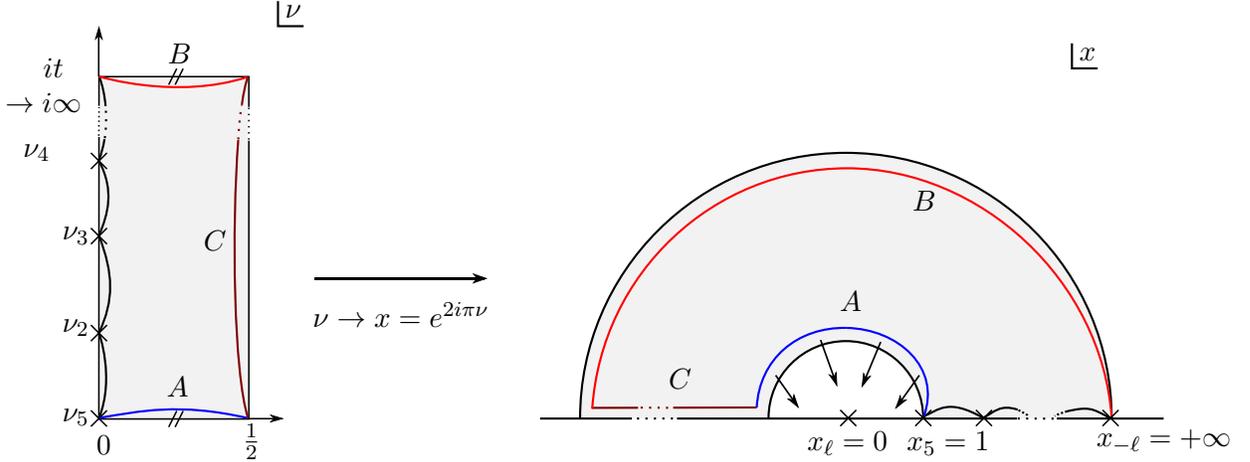}
  \caption{Mapping of the annulus to the plane with $x=e^{2i\pi \nu}$ in
    the $t\to \infty$ limit. In the reduction of the
    blue contour onto the real axis, a singularity at $x=0$ is encountered.}
\label{fig:map}
\end{figure}

The derivation of the monodromy relations in~\cite{Tourkine:2016bak}
was based on applying Cauchy's theorem to the open-string integrand
in a loop momentum representation (crucial for holomorphy).
We shall use the example of a five-point amplitude where $1$ is
circulated around the worldsheet boundary. The contour we study is
pictured in the left-hand side of figure~\ref{fig:map}.

At this point, strictly speaking we depart from the computation
of~\cite{Tourkine:2016bak}. We will follow the same reasoning, but
applied to the single-cut of the one-loop amplitude, instead of the
full string amplitude.
These were defined in \sec{sec:single-cut} by the
insertion of a $\delta^{(+)}(\ell ^2)$ operator. We proved there that
this induces the annulus to become an infinitely long strip.
Via the exponential map $\nu\mapsto x=\exp{(2i\pi \nu)}$, we send the
strip to a portion of the upper-half plane, as shown in \fig{fig:map}.
The segments of the integration contour are therefore also mapped on the
upper-half plane.

The substance of the original monodromy-relations came from the
vertical integration contours.  In the present five-point planar
example in \eqn{e:MonString}, the $\Ree(\nu)=0$ contour gives rise to
$(p_i\cdot p_j)$-type phases in contrast with the $\Ree(\nu)=1/2$
contour $C$, which gives phases dependent on the loop momentum.
However, since we now consider the monodromies of the single cut
integrand, i.e. with the $\delta^{(+)}(\ell ^2)$ function inserted,
the boundary terms have a different fate than
in~\cite{Tourkine:2016bak}.

\subsubsection{String-theoretic boundary terms}
\label{sec:stringy-origin}

The original string theory monodromies were based on discarding the
$A$ and $B$ contours. This was justified because
the two integrals only differ by a shift in the loop
momentum. Namely, denoting $\mathcal{I}(\ell)$ the integrand of the
string amplitude $\mathcal{A}$ being integrated along the closed
contour, it was shown in~\cite{Tourkine:2016bak} that
\begin{equation}
  \label{eq:shift-string-monodromies}
  \mathcal{I}(\ell)\big|_{A} = \mathcal{I}(\ell+p_1)\big|_{B} .
\end{equation}
These terms vanish
after integration over the loop momentum. We explicitly checked
in~\cite{Tourkine:2016bak} for $\mathcal{N}=4$ super-Yang-Mills that
the numerators produced by the loop-momentum pre-factors
in~\eqn{e:MonQFT} cancel box propagators pairwise and produce six 
triangles, which cancel pairwise after a shift of the loop
momentum.

Disregarding this cancellation, the application of Cauchy's theorem on
the closed contour of \fig{fig:map} actually says that the left-hand
side of \eqn{e:MonQFT} is given by the difference of two terms that
differ only by a shift in the loop momentum.  That property obviously
descends to the field-theory limit $\alpha'\to 0$. We study the
consequence of this fact later in \sec{sec:role-boundary-terms}.

Note that the loop momentum arising in
the string amplitude in~\eqref{e:cAdef} has a global definition, given
by the integral along the $a$-cycle~\cite{DHoker:1988pdl}
\begin{equation}\label{e:loopDef}
  \ell_a :=\oint_a \partial X= \int_0^{1\over2} {\partial X\over\partial\nu} d\nu ,
\end{equation}
with a similar definition at higher genus order.  As was emphasised
in~\cite{Tourkine:2016bak}, this means the string-theory limit induces
a global definition of the loop momentum across all the Feynman graphs
produced in this limit. This phenomenon is also observed in
ambitwistor-string 
constructions~\cite{Geyer:2015jch,Geyer:2015bja,Geyer:2016wjx}.

\subsubsection{Recovering the single-cut monodromy relation}

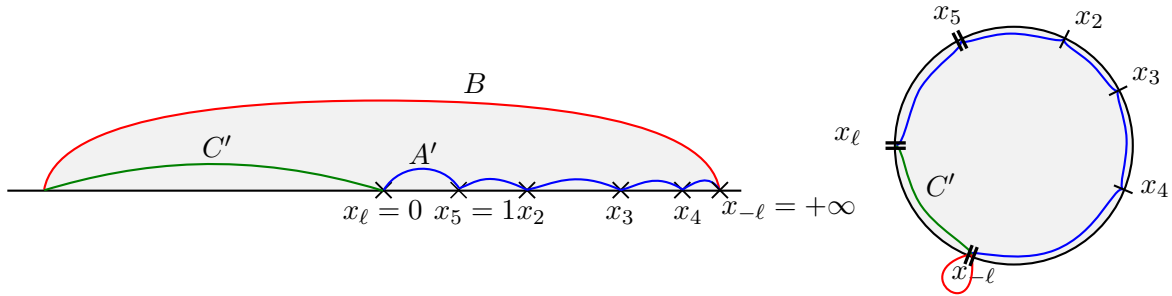
\begin{figure}[t]
  \centering
  \definecolor{cf2f2f2}{RGB}{242,242,242}
\definecolor{cff0000}{RGB}{255,0,0}
\definecolor{c0000ff}{RGB}{0,0,255}
\definecolor{c008000}{RGB}{0,128,0}

\begin{tikzpicture}[y=0.80pt, x=0.80pt, yscale=-1.000000, xscale=1.000000, inner sep=0pt, outer sep=0pt]
  \path[draw=black,fill=cf2f2f2,miter limit=4.00,nonzero rule,line width=0.800pt]
    (526.2943,175.1805) circle (1.5815cm);
  \path[draw=cff0000,fill=cf2f2f2,line join=miter,line cap=butt,even odd rule,line
    width=0.800pt] (67.8111,196.0131) .. controls (73.5126,163.8269) and
    (145.9913,153.8411) .. (226.7568,153.8411) .. controls (307.5223,153.8411) and
    (381.2179,163.5434) .. (387.0348,195.7297);
  \path[draw=black,line join=miter,line cap=butt,even odd rule,line width=0.800pt]
    (50.3950,196.4801) -- (397.4394,196.4801);
  \begin{scope}[cm={{0.80315,0.0,0.0,0.80315,(-217.35095,-33.84517)}},line width=0.996pt]
    \begin{scope}[line width=0.996pt]
      \path[draw=black,line join=miter,line cap=butt,even odd rule,line width=0.719pt]
        (634.5916,291.1870) -- (644.0486,281.7300);
      \path[draw=black,line join=miter,line cap=butt,even odd rule,line width=0.719pt]
        (644.0486,291.1870) -- (634.5916,281.7300);
    \end{scope}
  \end{scope}
  \path[draw=black,line join=miter,line cap=butt,even odd rule,line width=0.577pt]
    (259.9719,200.0219) -- (267.5674,192.4264);
  \path[draw=black,line join=miter,line cap=butt,even odd rule,line width=0.577pt]
    (267.5674,200.0219) -- (259.9719,192.4264);
  \begin{scope}[cm={{0.80315,0.0,0.0,0.80315,(-225.38246,-33.84517)}},line width=0.996pt]
    \begin{scope}[line width=0.996pt]
      \path[draw=black,line join=miter,line cap=butt,even odd rule,line width=0.719pt]
        (736.0266,291.1871) -- (745.4837,281.7300);
    \end{scope}
    \path[draw=black,line join=miter,line cap=butt,even odd rule,line width=0.719pt]
      (745.4837,291.1871) -- (736.0266,281.7300);
  \end{scope}
  \begin{scope}[cm={{0.80315,0.0,0.0,0.80315,(-218.95725,-33.84517)}},line width=0.996pt]
    \path[draw=black,line join=miter,line cap=butt,even odd rule,line width=0.719pt]
      (691.7606,291.1870) -- (701.2177,281.7300);
    \path[draw=black,line join=miter,line cap=butt,even odd rule,line width=0.719pt]
      (701.2177,291.1870) -- (691.7606,281.7300);
  \end{scope}
  \begin{scope}[cm={{0.80315,0.0,0.0,0.80315,(-285.17233,-33.84517)}},line width=0.996pt]
    \path[draw=black,line join=miter,line cap=butt,even odd rule,line width=0.719pt]
      (634.5916,291.1870) -- (644.0486,281.7300);
    \path[draw=black,line join=miter,line cap=butt,even odd rule,line width=0.719pt]
      (644.0486,291.1870) -- (634.5916,281.7300);
  \end{scope}
  \path[draw=c0000ff,line join=miter,line cap=butt,even odd rule,line
    width=0.800pt] (387.0348,195.8911) .. controls (382.2911,188.7543) and
    (376.0734,191.9353) .. (369.7351,195.9607);
  \path[draw=c0000ff,line join=miter,line cap=butt,even odd rule,line
    width=0.800pt] (369.7351,195.9608) .. controls (356.8207,187.7425) and
    (349.3528,193.8213) .. (340.4834,196.2214);
  \path[draw=c0000ff,line join=miter,line cap=butt,even odd rule,line
    width=0.800pt] (340.4834,196.2214) .. controls (318.5891,186.3401) and
    (308.8296,193.3127) .. (296.0580,196.1018);
  \path[draw=c0000ff,line join=miter,line cap=butt,even odd rule,line
    width=0.800pt] (296.0580,196.1018) .. controls (276.9285,186.6370) and
    (271.9035,192.5590) .. (264.0481,195.3933);
  \path[draw=c0000ff,line join=miter,line cap=butt,even odd rule,line
    width=0.800pt] (264.0481,195.3932) .. controls (255.8514,182.6710) and
    (236.3405,183.1548) .. (228.4245,196.1090);
  \path[draw=c008000,line join=miter,line cap=butt,even odd rule,line
    width=0.800pt] (228.0188,196.5147) .. controls (162.9873,176.7218) and
    (112.2278,182.9605) .. (67.8111,196.2968);
  \begin{scope}[cm={{0.80315,0.0,0.0,0.80315,(-126.14851,-33.84517)}},line width=0.996pt]
    \path[draw=black,line join=miter,line cap=butt,even odd rule,line width=0.719pt]
      (634.5916,291.1870) -- (644.0486,281.7300);
    \path[draw=black,line join=miter,line cap=butt,even odd rule,line width=0.719pt]
      (644.0486,291.1870) -- (634.5916,281.7300);
  \end{scope}
  \path[fill=black,line join=miter,line cap=butt,line width=0.800pt]
    (240.1392,183.9037) node[above right] (text7637) {$A'$};
  \path[fill=black,line join=miter,line cap=butt,line width=0.800pt]
    (265.8499,150.3665) node[above right] (text7649-1) {$B$};
  \path[fill=black,line join=miter,line cap=butt,line width=0.800pt]
    (142.5316,180.5876) node[above right] (text7649-5) {$C'$};
  \path[fill=black,line join=miter,line cap=butt,line width=0.800pt]
    (208.9685,212.0669) node[above right] (text4616) {$x_\ell=0$};
  \path[fill=black,line join=miter,line cap=butt,line width=0.800pt]
    (291.2700,212.0669) node[above right] (text4620) {$x_2$};
  \path[fill=black,line join=miter,line cap=butt,line width=0.800pt]
    (333.2350,212.0669) node[above right] (text4624) {$x_3$};
  \path[fill=black,line join=miter,line cap=butt,line width=0.800pt]
    (252.1203,212.0669) node[above right] (text4628) {$x_5=1$};
  \path[fill=black,line join=miter,line cap=butt,line width=0.800pt]
    (365.3610,212.0669) node[above right] (text4632) {$x_4$};
  \path[xscale=0.735,yscale=1.361,fill=black,line join=miter,line cap=butt,line
    width=0.800pt] (528.0193,155.0050) node[above right] (text4636)
    {$x_{-\ell}=+\infty$};
  \path[fill=black,line join=miter,line cap=butt,line width=0.800pt]
    (488.1547,117.5286) node[above right] (text4628-6) {$x_5$};
  \path[fill=black,line join=miter,line cap=butt,line width=0.800pt]
    (441.3959,174.8408) node[above right] (text4616-1) {$x_\ell$};
  \path[xscale=0.735,yscale=1.361,fill=black,line join=miter,line cap=butt,line
    width=0.800pt] (675.8871,177.3233) node[above right] (text4636-9)
    {$x_{-\ell}$};
  \path[fill=black,line join=miter,line cap=butt,line width=0.800pt]
    (555.1724,119.7278) node[above right] (text4620-1) {$x_2$};
  \path[fill=black,line join=miter,line cap=butt,line width=0.800pt]
    (582.3170,146.0149) node[above right] (text4624-4) {$x_3$};
  \path[fill=black,line join=miter,line cap=butt,line width=0.800pt]
    (586.2483,199.7063) node[above right] (text4632-8) {$x_4$};
  \path[draw=c0000ff,line join=miter,line cap=butt,even odd rule,line
    width=0.800pt] (506.1240,226.4915) .. controls (504.3757,223.8612) and
    (527.9231,231.6533) .. (545.7456,224.5844) .. controls (566.3145,216.4263) and
    (573.2046,194.2357) .. (578.1393,195.8175);
  \path[draw=c0000ff,line join=miter,line cap=butt,even odd rule,line
    width=0.800pt] (578.0415,195.9153) .. controls (575.4371,194.5470) and
    (579.9564,183.5639) .. (579.5982,172.1609) .. controls (579.2729,161.8034) and
    (573.6780,150.3056) .. (575.8211,149.2258);
  \path[draw=c0000ff,line join=miter,line cap=butt,even odd rule,line
    width=0.800pt] (575.9002,149.1752) .. controls (572.9418,150.3188) and
    (569.1350,142.6032) .. (563.1805,136.7204) .. controls (557.6339,131.2406) and
    (549.1036,127.3300) .. (550.2187,124.8616);
  \path[draw=c0000ff,line join=miter,line cap=butt,even odd rule,line
    width=0.800pt] (550.3367,124.5085) .. controls (548.8682,127.3366) and
    (536.7865,122.0866) .. (524.6854,122.1738) .. controls (513.4879,122.2545) and
    (502.1989,127.8613) .. (501.0823,125.7768);
  \path[draw=c0000ff,line join=miter,line cap=butt,even odd rule,line
    width=0.800pt] (470.7027,174.9576) .. controls (474.6615,174.9234) and
    (477.2020,155.2111) .. (482.4616,147.4427) .. controls (490.4165,135.6933) and
    (501.9071,129.7742) .. (501.1715,126.0842);
  \path[draw=black,line join=miter,line cap=butt,even odd rule,line width=0.800pt]
    (552.3968,120.5302) -- (548.0815,129.0657);
  \path[draw=black,line join=miter,line cap=butt,even odd rule,line width=0.800pt]
    (579.8891,147.0842) -- (571.3226,151.3376);
  \path[draw=black,line join=miter,line cap=butt,even odd rule,line width=0.800pt]
    (582.4275,197.7266) -- (573.6280,193.9787);
  \path[draw=cff0000,line join=miter,line cap=butt,even odd rule,line
    width=0.800pt] (506.2992,226.3974) .. controls (510.6060,263.7519) and
    (473.1583,234.7142) .. (506.2037,226.4318);
  \path[draw=c008000,line join=miter,line cap=butt,even odd rule,line
    width=0.800pt] (506.2101,226.6760) .. controls (506.2800,223.1616) and
    (488.7503,214.0767) .. (481.6242,202.2130) .. controls (474.5884,190.4996) and
    (474.0406,176.1192) .. (470.5472,175.4466);
  \begin{scope}[cm={{0.23008,-0.59466,0.59466,0.23008,(125.28556,447.3415)}},line width=0.996pt]
    \path[draw=black,line join=miter,line cap=butt,even odd rule,line width=1.255pt]
      (530.0000,434.3622) -- (545.0000,434.3622);
    \path[draw=black,line join=miter,line cap=butt,even odd rule,line width=1.255pt]
      (530.0000,430.3622) -- (545.0000,430.3622);
  \end{scope}
  \begin{scope}[cm={{0.63762,0.0,0.0,0.63762,(127.68912,-100.52865)}},line width=0.996pt]
    \path[draw=black,line join=miter,line cap=butt,even odd rule,line width=1.255pt]
      (530.0000,434.3622) -- (545.0000,434.3622);
    \path[draw=black,line join=miter,line cap=butt,even odd rule,line width=1.255pt]
      (530.0000,430.3622) -- (545.0000,430.3622);
  \end{scope}
  \begin{scope}[cm={{0.2904,0.56765,-0.56765,0.2904,(590.13978,-305.18191)}},line width=0.996pt]
    \path[draw=black,line join=miter,line cap=butt,even odd rule,line width=1.255pt]
      (530.0000,434.3622) -- (545.0000,434.3622);
    \path[draw=black,line join=miter,line cap=butt,even odd rule,line width=1.255pt]
      (530.0000,430.3622) -- (545.0000,430.3622);
  \end{scope}
  \path[fill=black,line join=miter,line cap=butt,line width=0.800pt]
    (484.7054,198.8904) node[above right] (text7649-5-3) {$C'$};

\end{tikzpicture}
\kern-1cm 
  \caption{The right-hand side integration contour of \Fig{fig:map} is
  equivalent to the one on the left-hand side here, which is equivalently
represented in a disk picture on the right.}
 \label{fig:map2}
\end{figure}

To finish connecting the two monodromy relations~\eqref{e:FundBCJF1}
and~\eqref{e:MonQFT},
we need to deform the contour of~\fig{fig:map} down to the real axis.
No singularity is present at $x=-1$, while one is present at $0$
and the contours $A$ and $C$ can be turned into the
contours $A'$ and $C'$ of \fig{fig:map2}.
The contour $B$ vanishes as a consequence of onshellness and momentum conservation
(which guarantees $SL(2,\mathbb R)$-invariance of the whole integral).
This actually tells us that we are truly on the projective
line and that the contour $C'$ connects the puncture at $x=0$ and
$x=+\infty$ as shown in the disk representation on the right-hand side
of \fig{fig:map}.

Notice that two additional states with momentum $\pm \ell$ appear at
$x=0,+\infty$. This intuitive fact is easily derived by inspection of the degeneration of the
one-loop Koba-Nielsen factor of string theory when $t\to0$.  In
essence, such a computation is exposed
in~\cite{Geyer:2015jch,Geyer:2015bja,Geyer:2016wjx}.

What we have shown so far is that, starting from a five-point
amplitude (the example used here), we obtained five blue contours in
\fig{fig:map2} that correspond to the five terms from the circulation
of the vertex operator 1 in the fundamental monodromy
relation~\eqref{e:FundBCJF1}.  However, the one-loop monodromy
relation in~\eqref{e:MonQFT} had only four terms. The resolution of
this apparent contradiction goes as follows.

Our starting point was actually not the original string-theory
monodromies, but the monodromies of the projected
amplitude~\eqref{e:cAproj} with the insertion of the single-cut
operator $\delta^{(+)}(\ell^2)$.  This term does not affect the
phases, but it does break the freedom to shift the loop momentum to
cancel out the integral between the contours $A$ and $B$.
Adding the onshellness and momentum conservation condition, we see
that the contribution from the contour $B$ drops out, and the contour $A$ gives an
additional term that connects $x=0$ to $x=1$.
One can then check that the phases of the momentum kernel,
and the different terms match.

This concludes the proof of the correspondence between the
string-theory monodromies and the forward monodromies.

\subsection{Dissecting the field theory limit of the string theory monodromies}
\label{sec:dissecting}

In this section we present two refinements of the string-theory
monodromies. First, we comment on the inner structure of planar and
non-planar diagrams arising from the string theory computation. Then we
elaborate on the role of boundary terms from the string-theoretic
perspective. 
This will allow us to describe the
BCJ-compatibility of the string-theory monodromies in \sec{sec:fourpoints}.

To our knowledge these results are new and should be
seen as an important refinement of the field-theory monodromies
originally discovered in~\cite{Boels:2011tp,Boels:2011mn}.

\subsubsection{Role of boundary terms}
\label{sec:role-boundary-terms}

Let us come back to the interpretation of the string-theory monodromies
in the light of the observation made in \sec{sec:stringy-origin}.
What we explained in eq.~(\ref{eq:shift-string-monodromies}) and below
was  that  the terms  in  the  right-hand  side of  the  string-theory
monodromies needed to be of the form $F(\ell+p_1)-F(\ell )$.

Here we would like to motivate the following conjecture: the role
of these terms is to accomodate for the fact that when the so-called
``BCJ-moves'' are done around a loop, a shift of the loop momentum
needs to be made in the last move. 
This conjecture is motivated by the graphical argument that
follows, and by the explicit examples in field theory at four and
five points presented in \sec{sec:fourpoints}.

\begin{figure}[t]
  \centering
  \footnotesize
  \input{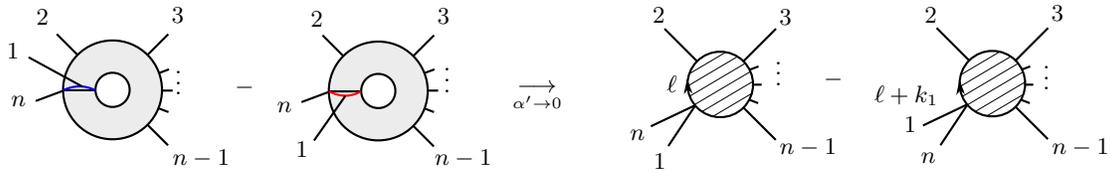}
  \caption{Left-hand side: boundary terms along contours $A$ and
    $B$ as defined in \fig{fig:map}. Right-hand side:
    conjectured corresponding types of one-loop graphs.
    Notice the loop-momentum shift.}\label{fig:pinching-ftl}
\end{figure}

From before, we know that the terms that integrate to zero in the
string theory monodromies come from the $A$ and $B$ contour integrals
in \fig{fig:map}. In the field-theory limit, we conjecture that these
graphs need to appear as pinched contributions. Intuitively the
reasoning is clear, and exposed in \fig{fig:pinching-ftl}. The drawing
shows that the difference between the $A$ and $B$ contours results in
graphs with a pinched propagator, that only differ by a shift of the
loop momentum. In string theory, the shift arises because the loop
momentum jumps when a puncture goes through the $a$-cycle on which it
is defined, see~\eqn{e:loopDef}.

An actual extraction of the field theory limit to all loops, in the spirit of the ``string-based rules''
\cite{Bern:1990cu,Bern:1990ux,Bern:1991aq,Bern:1993mq,Schubert:2001he},
would be interesting but very technical in nature and is outside of
the scope of this paper.
Such a computation should however prove two properties. 

First, that these graphs, coming from the $A$ and $B$ contours, appear
at order $\alpha'$. This is required by the fact that the integrand
monodromy relations coming from string theory come at order $\mathcal{O}(\alpha')$.
On dimensional grounds, it is clear that such pinched-terms contribute
at order in $\alpha'$: the pinching of $1$ and $n$ removes a
propagator $1/(\alpha' p^2)$, therefore it contributes at order
$\alpha'$, even though it is a purely field-theoretic contribution.

Relatedly, the second part of the computation is to check that no 'triangle'-type
graphs are generated in the process (i.e. graphs where a propagator of
the form $1/(p_1\cdot p_n)$. This is important because the momentum
factor in the denominator has an inverse power of $\alpha'$
that would make these terms contribute at~${\cal O}(1)$. Of course,
because of the generic functional form $F(\ell+p_1)-F(\ell)$, these
terms would still cancel after integration, even if present.

\subsubsection{Planar versus non-planar refinement}
\label{sec:planar-vs-non}

In this section we clarify the connection of the string-theory
monodromy relations~\eqref{e:MonQFT} to the colour-kinematic
duality. The result we present is that they can be rewritten as
\begin{equation}
  \label{eq:monodromy-refined}
   I^\flat(1,2,\ldots,n)
 + \sum_{i=2}^{n-1} \Big[ (\ell \cdot p_1) I^\flat(2,\ldots,i,1,i+1,\ldots,n)
    + (p_1\cdot p_{2 \ldots i})  I(2,\ldots,i,1,i+1,\ldots,n) \Big] \approx 0 ,
\end{equation}
where the $\approx$ sign indicates
again that the relation is valid modulo terms of the form
$F(\ell+p_1)-F(\ell)$ that integrate to zero.  Let us explain this new
form of the monodromy relations.

The ``flattened'' integrand $I^{\flat}$ is defined as
follows.
Start from the non-planar one-loop string integrand
$\mathcal{I}(2,\ldots,n;1)$ with leg $1$ on one boundary and
the other states on the other boundary.
The traditional way of obtaining the integrand monodromies is to use
the $U(1)$-decoupling relations of~\cite{Bern:1994zx} to rewrite this
non-planar contribution as a sum of planar contributions. The outcome
are the integrand relations we talked about that involve only planar integrands.

Instead of doing this, we will use the antisymmetry of the cubic
vertices of the graphs entering the non-planar integrand in order to
write them in a planar fashion. The sum of these graphs, for a
particular ordering $(2,\dots,i,1,i+1,n)$, constitutes the term we
call $I^\flat(2,\dots,i,1,i+1,n)$.  The existence of such a
representation, involving only graphs with antisymmetric cubic
vertices, can be justified using the string-based
rules~\cite{Bern:1990cu,Bern:1990ux,Bern:1991aq,Bern:1993mq,Schubert:2001he}. From
the string-theoretic perspective, the idea is that the integrand can
always undergo a succession of integrations by parts which allows to
reduce it to a sum of only trivalent graphs. (For more details on
this, see the review of the worldline
formalism~\cite{Schubert:2001he}.)  This is related to the fact that
in string theory, gauge invariance is enforced differently than with
the usual Feynman rules and does not require contact terms but only
BRST closure of the expressions~\cite{Bjornsson:2010wu}.

\begin{figure}[t]
   \centering
\def\one{$\color{blue} 1$}
\input{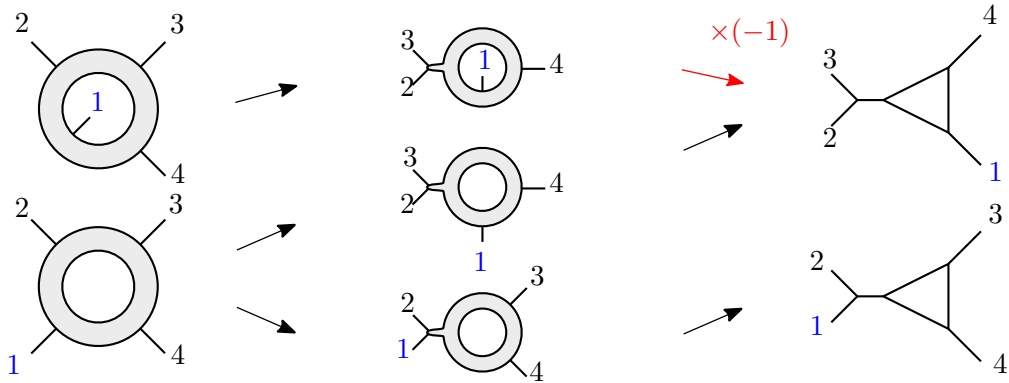}
\caption{Triangles obtained by bringing $\nu_2$
   close to $\nu_3$, or to $\nu_1$.
   The latter is not possible for the upper vertex configuration,
   in which $\nu_1$ is on the other boundary of the annulus.}
\label{fig:FFT}
\end{figure}

The peculiarity of the graphs entering $I^\flat(2,\dots,i,1,i+1,n)$
is that the leg $1$, being attached to the other colour trace,
can never belong to an external tree attached to the loop.
This is obvious from the string-theoretic perspective as well:
the regions of integration
giving rise to such trees are those for which a pair of punctures
$\nu_i$ and $\nu_j$ become infinitesimally close.  This can not happen
for the leg $1$, which is always on the opposite side of the annulus.
This reasoning is illustrated in \fig{fig:FFT}.

This definition implies that 
\begin{equation}
  \label{eq:Iflat-def}
  \lim_{\alpha'\to0} \mathcal{I}(2,
\dots,n;1)=-\sum_{i=1}^n I^\flat(p_2,\dots,\!\!\!
\underbrace{p_1}_{\textrm{position}~i}\!\!\!,\dots,p_{n}) ,
\end{equation}
where it is understood that $I^\flat$ has no subtrees of the form of
the lower one on the right-most side of \fig{fig:FFT}. Since there is
only one vertex to flip, this only results in a global sign.

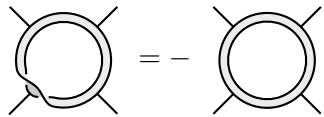
\begin{figure}[t]
   \centering \definecolor{cececec}{RGB}{236,236,236}
\definecolor{ccccccc}{RGB}{204,204,204}
\definecolor{cffffff}{RGB}{255,255,255}

\begin{tikzpicture}[y=0.80pt, x=0.80pt, yscale=-1.000000, xscale=1.000000, inner sep=0pt, outer sep=0pt]
  \path[draw=cececec,fill=cececec,miter limit=4.00,nonzero rule,line
    width=0.800pt] (173.4666,176.7983) .. controls (171.7611,176.4994) and
    (171.0022,176.2012) .. (170.1938,175.5124) .. controls (169.6379,175.0388) and
    (168.7681,173.9222) .. (168.1854,172.9342) .. controls (168.0330,172.6757) and
    (171.5426,172.8131) .. (173.4666,173.1410) .. controls (178.8798,174.0634) and
    (183.9874,172.9315) .. (188.0861,169.9014) .. controls (197.0340,163.2862) and
    (198.2215,150.5575) .. (190.6501,142.4180) .. controls (188.4245,140.0255) and
    (185.0119,137.9970) .. (181.8145,137.1661) .. controls (179.4715,136.5573) and
    (175.8995,136.4549) .. (173.5870,136.9302) .. controls (169.7115,137.7269) and
    (166.7753,139.2726) .. (164.0679,141.9416) .. controls (159.7237,146.2240) and
    (157.9157,151.9498) .. (158.8598,158.4357) .. controls (159.0912,160.0257) and
    (159.1883,163.6291) .. (158.9997,163.6291) .. controls (158.7795,163.6291) and
    (157.3382,162.4723) .. (156.7005,161.7837) .. controls (155.3859,160.3640) and
    (155.0165,158.8693) .. (155.0241,155.0000) .. controls (155.0301,152.0486) and
    (155.3285,150.2384) .. (156.2284,147.6991) .. controls (157.9714,142.7812) and
    (161.5820,138.4618) .. (166.0850,135.9074) .. controls (179.2380,128.4464) and
    (195.6691,135.8550) .. (198.8602,150.6855) .. controls (199.2294,152.4015) and
    (199.2710,157.3581) .. (198.9307,159.0749) .. controls (197.1086,168.2643) and
    (189.7903,175.3226) .. (180.5350,176.8168) .. controls (178.7335,177.1077) and
    (175.1784,177.0984) .. (173.4666,176.7983) -- cycle;
  \path[fill=ccccccc,miter limit=4.00,nonzero rule,line width=0.800pt]
    (164.2931,172.1510) .. controls (162.9094,171.8512) and (161.5386,171.1106) ..
    (160.9047,170.3204) .. controls (160.6333,169.9821) and (160.2501,169.2480) ..
    (160.0531,168.6889) .. controls (159.6817,167.6346) and (159.2398,164.8219) ..
    (159.4177,164.6444) .. controls (159.5641,164.4984) and (162.8799,166.7011) ..
    (163.8328,167.5773) .. controls (164.7239,168.3968) and (165.8268,169.8788) ..
    (166.6880,171.4142) -- (167.2494,172.4151) -- (166.3167,172.4013) .. controls
    (165.8037,172.3943) and (164.8931,172.2811) .. (164.2931,172.1511) -- cycle;
  \path[draw=black,miter limit=4.00,nonzero rule,line width=0.800pt]
    (159.5028,166.6365) .. controls (159.6400,168.7499) and (160.0136,169.5262) ..
    (161.2556,170.7625) .. controls (162.3340,171.8359) and (163.3077,172.2392) ..
    (165.1617,172.3986)(170.0690,172.4772) .. controls (171.7198,172.4962) and
    (172.0125,172.5262) .. (173.2243,172.7881) .. controls (179.0909,174.0562) and
    (185.4638,172.4206) .. (190.0242,167.8811) .. controls (197.1534,160.7847) and
    (197.1534,149.2791) .. (190.0242,142.1827) .. controls (182.8950,135.0863) and
    (171.3363,135.0863) .. (164.2070,142.1827) .. controls (159.7866,146.5828) and
    (158.1070,152.6781) .. (159.1684,158.3663) .. controls (159.4088,159.6551) and
    (159.4181,159.9278) .. (159.4313,161.6468)(161.2233,139.2127) .. controls
    (155.7308,144.6800) and (153.6754,152.2731) .. (155.0572,159.3314) .. controls
    (155.8835,163.5522) and (161.1173,164.4744) .. (164.4018,167.7438) .. controls
    (167.5213,170.8489) and (168.1511,175.9577) .. (172.1586,176.8557) .. controls
    (179.4268,178.4842) and (187.3503,176.4827) .. (193.0079,170.8511) .. controls
    (201.7850,162.1144) and (201.7850,147.9494) .. (193.0079,139.2127) .. controls
    (184.2309,130.4760) and (170.0004,130.4760) .. (161.2233,139.2127) -- cycle;
  \path[draw=black,line join=miter,line cap=butt,even odd rule,line width=0.800pt]
    (151.3661,180.6630) -- (161.2548,170.8198);
  \path[draw=black,line join=miter,line cap=butt,even odd rule,line width=0.800pt]
    (161.2234,139.2127) -- (151.6880,129.7212);
  \path[draw=black,line join=miter,line cap=butt,even odd rule,line width=0.800pt]
    (193.0079,139.2127) -- (202.5433,129.7212);
  \path[draw=black,line join=miter,line cap=butt,even odd rule,line width=0.800pt]
    (193.0079,170.8511) -- (202.5433,180.3426);
  \path[cm={{0.70873,-0.70547,0.70873,0.70547,(0.0,0.0)}},draw=black,fill=cececec,miter
    limit=4.00,nonzero rule,line width=0.800pt] (82.9675,302.7560) circle
    (0.6328cm);
  \path[draw=black,line join=miter,line cap=butt,even odd rule,line width=0.800pt]
    (298.6899,180.2530) -- (289.2680,170.8745);
  \path[draw=black,line join=miter,line cap=butt,even odd rule,line width=0.800pt]
    (289.2680,139.2361) -- (298.8034,129.7446);
  \path[draw=black,line join=miter,line cap=butt,even odd rule,line width=0.800pt]
    (257.4835,139.2361) -- (247.9481,129.7446);
  \path[draw=black,line join=miter,line cap=butt,even odd rule,line width=0.800pt]
    (257.4835,170.8745) -- (247.9481,180.3660);
  \path[cm={{0.70873,-0.70547,0.70873,0.70547,(0.0,0.0)}},draw=black,fill=cffffff,miter
    limit=4.00,nonzero rule,line width=0.800pt] (82.9675,302.7560) circle
    (0.5175cm);
  \path[xscale=1.002,yscale=0.998,fill=black,line join=miter,line cap=butt,line
    width=0.800pt] (212.0587,159.3298) node[above right] (text4578) {$=-$};

\end{tikzpicture}
   \vspace{-7pt}
   \caption{The antisymmetry of the three-point vertex
      converts non-planar contributions to planar.}
   \label{fig:antisym}
 \end{figure}

Let us denote by $I(\sigma)$ the field-theory integrands coming from
planar contributions, these with $(p_i \cdot p_j)$-type phases in
the string-theory version
and hence $(p_i \cdot p_j)$ factors in the field-theory limit.
Similarly, we call $I^\flat(\sigma)$ the integrands
coming from non-planar contributions but made planar using the
antisymmetry of the three-point vertex for leg $1$, as in
\fig{fig:antisym}. The important point we want to make is that these
receive only phases of the form  $(\ell \cdot p_i)$,
or similar factors in the field-theory limit.
Together with the previous considerations, this justifies
the refined monodromy relation~\eqref{eq:monodromy-refined}.

Below we dissect this relation and the role the boundary terms
(which vanish after integration) in the case of a generic four-point
amplitude.

\subsection{BCJ-compatibility of the string theory monodromies}
\label{sec:fourpoints}

It was already observed in~\cite{Boels:2011tp} that the integrand
relations are satisfied at four points by $\mathcal{N}=4$
super-Yang-Mills. Shortly after~\cite{Tourkine:2016bak},
an $n$-point proof that
BCJ representations satisfy the string theory monodromies
was given in~\cite{Brown:2016hck,Brown:2016mrh}.
For the sake of illustrating the refinements that were made in
the previous section, we analyse the most general case of any massless
theory that can be obtained as a limit of open-string theory.
\subsubsection{Four-point case}
\label{sec:four-point-case}

Let us consider the refined monodromies~\eqref{eq:monodromy-refined} in
the four-point case:
\begin{equation}
\begin{aligned}
   (\ell \cdot p_1) I^\flat(1,2,3,4)
 + (\ell \cdot p_1) I^\flat(2,1,3,4) &
 + (\ell \cdot p_1) I^\flat(2,3,1,4) \\
 + (p_1 \cdot p_2) I(2,1,3,4) &
 + (p_1 \cdot p_{23}) I(2,3,1,4) \approx 0 .
\end{aligned}
\label{eq:monodromy-refined-4pt}
\end{equation}
We assume a cubic representation for the integrands composed of
boxes, triangles and massive bubbles, i.e. the four-point topologies
that do not integrate to zero in dimensional regularisation.
Note again that
$I^\flat(\sigma)$ lacks some of the diagrams present in $I(\sigma)$ ---
those with the special leg~1 in a massive corner of a triangle or a
bubble represented in the first line of~\eqref{eq:monodromy-refined-4pt2}.
We rewrite~\eqref{eq:monodromy-refined-4pt} by separating these
contributions from the integrands of $I(2,1,3,4)$ and $I(2,3,1,4)$  in the
second line and regroup the remaining pieces, $I^\flat(2,1,3,4)$ and
$I^\flat(2,3,1,4)$ with the corresponding ones from the first line
\begin{align}
\label{eq:monodromy-refined-4pt2}
   (\ell \cdot p_1) I^\flat(1,2,3,4)
 + ((\ell+p_2) \cdot p_1) I^\flat(2,1,3,4) &
 + ((\ell+p_{23}) \cdot p_1) I^\flat(2,3,1,4) \\
 + (p_1 \cdot p_2) \left[I(2,1,3,4)-I^\flat(2,1,3,4)\right] &
 + (p_1 \cdot p_{23}) \left[I(2,3,1,4)-I^\flat(2,3,1,4)\right] \approx 0 . \nn
\end{align}
We can rewrite the  kinematic coefficients in the first line as differences of propagators:
\begin{equation}
   2(\ell \cdot p_1) = (\ell+p_1)^2 - \ell^2\!, \quad
   2(\ell+p_2)\cdot p_1 = (\ell+p_{12})^2 - (\ell+p_{2})^2\!, \quad
   2(\ell+p_{23})\cdot p_1 = (\ell-p_{4})^2 - (\ell-p_{14})^2\! .
\end{equation}
The cubic expansion for the monodromy relation~\eqref{eq:monodromy-refined-4pt} is then
\small
\begin{align} 
 0 ~~~\approx~~~ s_{12}
 & I\!\left[ \scalegraph{0.45}{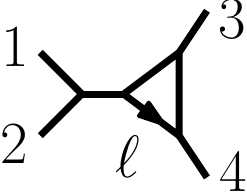}\!\!
        +\!\!\scalegraph{0.45}{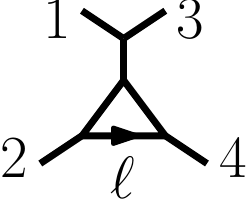}\!\!
          +\!\scalegraph{0.45}{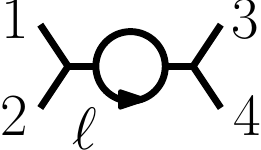}\!
           + \scalegraph{0.45}{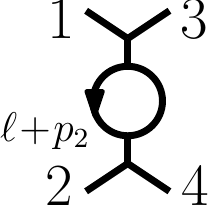}
      \right]
 - s_{14}
   I\!\left[ \scalegraph{0.45}{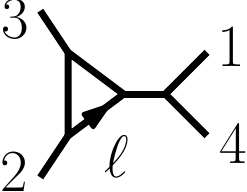}\!
        +\!\!\scalegraph{0.45}{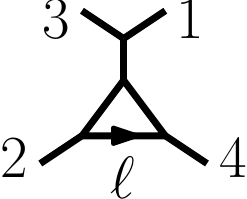}\!\!
          +\!\scalegraph{0.45}{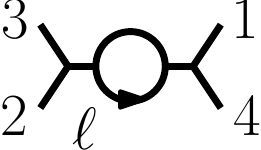}\!
           + \scalegraph{0.45}{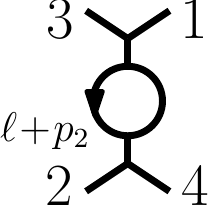}
      \right]\nn \!\!\!\!\!\!\!\\\!\!\!\!\!\!
 + [(\ell\!+\!p_1)^2\!- \ell^2]\:\!
 & I\!\left[ \scalegraph{0.45}{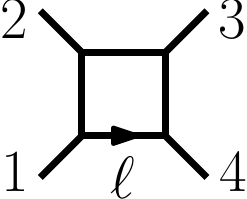}\!\!
          +\!\scalegraph{0.45}{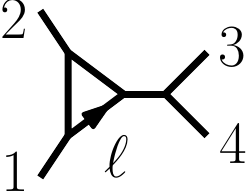}\!
           + \scalegraph{0.45}{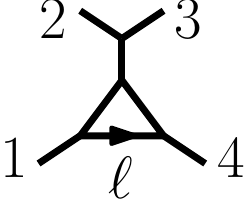}
      \right] 
 + [(\ell\!+\!p_{12})^2\!- (\ell\!+\!p_{2})^2]\:\!
   I\!\left[ \scalegraph{0.45}{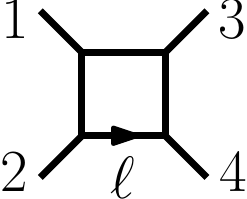}\!\!
          +\!\scalegraph{0.45}{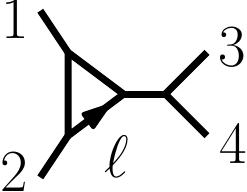}\!
           + \scalegraph{0.45}{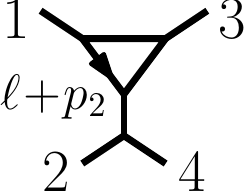}
      \right] \nn \!\!\!\!\!\!\!\\ & \qquad\qquad\qquad\qquad\quad~\:\,
 + [(\ell\!-\!p_{4})^2\!- (\ell\!-\!p_{14})^2]\:\!
   I\!\left[ \scalegraph{0.45}{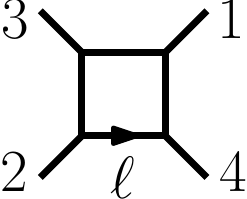}\!\!
          +\!\scalegraph{0.45}{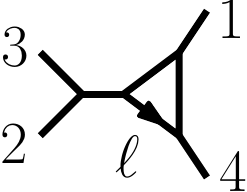}\!\!
           + \scalegraph{0.45}{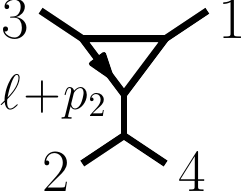}
      \right] .\!\!
\label{eq:monodromy-refined-4pt3}
\end{align}
\normalsize
Remember that, although \eqn{eq:monodromy-refined-4pt3}
may seem gauge-dependent,
its invariance is guaranteed
by the $U(1)$-decoupling identity~\cite[eq.~(6.4)]{Bern:1990ux},
\be
   I^\flat(1,2,3,4) + I^\flat(2,1,3,4) + I^\flat(2,3,1,4)
    = - I(2,3,4;1) .
\ee
Now we can collect terms with the same loop propagator structure.
Some of them turn out to correspond to massless bubbles,
so we omit them; the remaining ones are
\beal \!\!\!\!\!\!\!\!\!\!
 0 ~\approx~
   \frac{1}{\ell^2 (\ell\!+\!p_{12})^2 (\ell\!-\!p_4)^2} &\!
   \left\{ n\!\left(\scalegraph{0.45}{box1234}\right)
         - n\!\left(\scalegraph{0.45}{box2134}\right)
         + n\!\left(\scalegraph{0.45}{triangle2134sL}\right)\!
   \right\} \\ \!\!\!\!\!\!\!\!\!\!
 - \frac{1}{(\ell\!+\!p_1)^2 (\ell\!+\!p_{12})^2 (\ell\!-\!p_{4})^2}
           \!\!\!\!&\,\,\,\,\,
           n\!\left(\scalegraph{0.45}{box1234}\right)
 + \frac{1}{\ell^2 (\ell\!+\!p_2)^2 (\ell\!+\!p_{23})^2}\!
   \left\{ n\!\left(\scalegraph{0.45}{box2314}\right)
         - n\!\left(\scalegraph{0.45}{triangle2314tR}\right)\!
   \right\} \!\!\!\!\!\!\!\!\!\!\!\!\!\!\!\!\!\!\!\! \\
 + \frac{1}{\ell^2 (\ell\!+\!p_2)^2 (\ell\!-\!p_4)^2} &\!
   \left\{ n\!\left(\scalegraph{0.45}{box2134}\right)
         - n\!\left(\scalegraph{0.45}{box2314}\right)
         + n\!\left(\scalegraph{0.45}{triangle2134uT}\right)\!
   \right\} \\
 + \frac{1}{s_{12} \ell^2 (\ell\!+\!p_{12})^2} &\!
   \left\{ n\!\left(\scalegraph{0.45}{triangle1234sR}\right)
         - n\!\left(\scalegraph{0.45}{triangle2134sR}\right)
         + n\!\left(\scalegraph{0.45}{bubble2134s}\right)\!
   \right\} \\
 - \frac{1}{s_{23} (\ell\!+\!p_1)^2 (\ell\!-\!p_4)^2}
           \!\!\!\!&\,\,\,\,\,
           n\!\left(\scalegraph{0.45}{triangle1234tT}\right)
 + \frac{1}{s_{23} \ell^2 (\ell\!+\!p_{23})^2}\!
   \left\{ n\!\left(\scalegraph{0.45}{triangle2314tL}\right)
         - n\!\left(\scalegraph{0.45}{bubble2314t}\right)\!
   \right\} \!\!\!\!\!\!\!\!\!\! \\
 + \frac{1}{s_{24}(\ell\!+\!p_2)^2 (\ell\!-\!p_4)^2} &\!
   \left\{ n\!\left(\scalegraph{0.45}{triangle2134uB}\right)
         - n\!\left(\scalegraph{0.45}{triangle2314uB}\right)
         + n\!\left(\scalegraph{0.45}{bubble2134u}\right)\!
   \right\} . 
\label{eq:monodromy-refined-4pt4}
\eeal 
We immediately see that the cubic numerators organise themselves
into triplets related by kinematic Jacobi identities.  Note that in
the second and fifth lines, one should shift the loop-momentum in one
of the diagrams to reconstruct the correct Jacobi relations. 

This is completely consistent with the exact-integrand relationship
picture from string theory. In \sec{sec:loopvsforward} we explained that
the right-hand side of \eqn{eq:monodromy-refined-4pt} has to be composed
of terms of the form $F(\ell -p_1)-F(\ell )$,
if the integrand relationship was
written in a representation that is obtainable from the string-based
rules. Here we see that only the terms which involve a loop momentum
shift of the form $\ell\to\ell-p_1$ fail to constitute exact BCJ triplets.

From what we obtained in field theory, it seems that we can actually
go further and constrain the form of string representations that would
produce BCJ numerators. 

Let us assume that we have one, and write down the refined monodromy
relations, as in \eqref{eq:monodromy-refined-4pt4}. All the BCJ
triplets that do not involve a loop-momentum shift vanish, and we are
left with those which involve a shift. They ought to equal the terms
coming from the $A,B$ contours in the left-hand side; this constitutes
a universal constraint on string-theoretic representations of BCJ
numerators.

\subsubsection{Five-point case}
\label{sec:five-point-case}

At higher points, distinct loop propagator structures
appear with more than three cubic numerators.
For example, the residue structure of the one-mass triangle $(2,3,451)$
is shared by the following numerators:
\small
\begin{align}
 - \frac{1}{s_{45} (\ell+p_1)^2 (\ell\!+\!p_{12})^2 (\ell\!-\!p_{45})^2}
   n\!\left(\;\!\!\scalegraph{0.45}{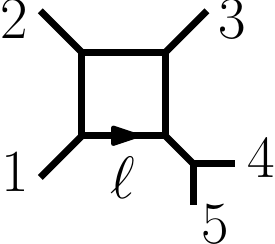}\!\!\right)
   \qquad\;\,\,& \nn \\
 + \frac{1}{\ell^2 (\ell\!+\!p_2)^2 (\ell\!+\!p_{23})^2}\!
   \left\{ \frac{1}{s_{45}} n\!\left(\;\!\!\scalegraph{0.45}{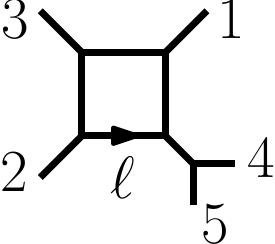}\!\!\right)
         + \frac{s_{12}\!+\!s_{13}}{s_{23}} \right.\:\!\!&\!
    \left[ \frac{1}{s_{45}}
           n\!\left(\;\!\!\scalegraph{0.45}{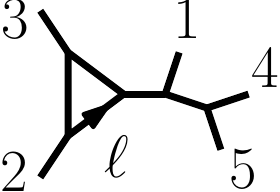}\!\!\right)
         + \frac{1}{s_{14}}
           n\!\left(\;\!\!\scalegraph{0.45}{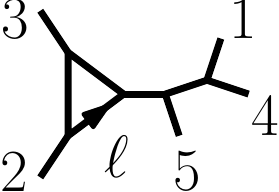}\!\!\right)
    \right] \nn \\
         + \frac{s_{12}\!+\!s_{13}\!+\!s_{14}}{s_{23}} &\:\!\!\!\left.
    \left[ \frac{1}{s_{15}}
           n\!\left(\;\!\!\scalegraph{0.45}{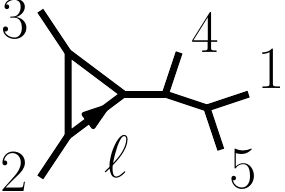}\!\!\right)
         + \frac{1}{s_{14}}
           n\!\left(\;\!\!\scalegraph{0.45}{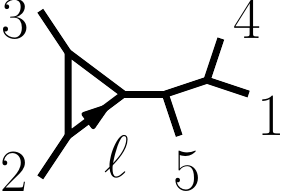}\!\!\right)
    \right]
   \right\} \\\!\!\!\!\!\!
 =\!-\frac{1}{s_{45} (\ell\!+\!p_1)^2 (\ell\!+\!p_{12})^2 (\ell\!-\!p_{45})^2}
           n\!\left(\;\!\!\scalegraph{0.45}{box12345}\!\!\right)\!
         + \,\,\,\,\,\,\,\,\,\,\,\,\,&\!\!\!\!\!\!\!\!\!\!\!\!\!
           \frac{1}{s_{45} \ell^2 (\ell\!+\!p_2)^2 (\ell\!+\!p_{23})^2}\!
   \left\{ n\!\left(\;\!\!\scalegraph{0.45}{box23145}\!\!\right)
         - n\!\left(\;\!\!\scalegraph{0.45}{triangle231-45tR}\!\right)\!
   \right\} \!\!\!\!\!\!\! \nn \\
 + \frac{1}{s_{23} \ell^2 (\ell\!+\!p_2)^2 (\ell\!+\!p_{23})^2}\!
   \left\{ n\!\left(\;\!\!\scalegraph{0.45}{triangle231-45tR}\!\!\right)
         + \right.\,\,\,\,\,\,\,&\!\!\!\!\!\!\! \left.\!
           n\!\left(\;\!\!\scalegraph{0.45}{triangle23-41-5tR}\!\!\right)
         - n\!\left(\;\!\!\scalegraph{0.45}{triangle234-15tR}\!\!\right)\!
   \right\} . \nn
\end{align}
\normalsize

We performed the full
five-point computation, and checked a few higher-point examples which
support this claim.
In these cases, we could verify that such expressions correspond
to linear combinations of Jacobi identities.

This is consistent with the analysis
of~\cite{Brown:2016hck,Brown:2016mrh} where it was shown that a colour-dual
representation at $n$ points always satisfies the string-theory
monodromies. However, no higher-point $n$-plets appeared 
in this analysis. This is possibly a consequence of their use of the
multiperipheral representation of the
colour factors of~\cite{DelDuca:1999rs}.
It would be interesting to understand this point in more detail.

\section{Discussion}
\label{sec:discussion}

In this work we have studied the kinematic monodromy relations on
single cuts of one-loop amplitudes in gauge theory. These relations
are derived from the monodromy relations on the regularised forward
tree amplitudes. We have explained that the regularisation does not
affect the consequences of the monodromy relations for the
irreducible one-loop numerators. We have checked that the monodromy
relations are always satisfied by colour-dual numerators, in agreement
with the analysis of~\cite{Brown:2016hck,Brown:2016mrh}.

The string-theoretic construction of~\cite{Tourkine:2016bak} and the
two-loop field-theory relations~\cite{Feng:2011fja} show that
higher-loop monodromy relations mix planar and non-planar sectors in
interesting new ways.  Indeed, the tree-level BCJ relations have
already been used for a two-loop integrand calculation at full colour
in~\cite{Badger:2015lda}.  Moreover, they have been seen to reduce the
number of independent one-loop coefficients
in~\cite{Chester:2016ojq,Primo:2016omk}.  We look forward to these
techniques converging to a systematic tool for multi-loop
computations.

We have explained that the string-theory monodromy relations are exact
relations, and their field-theory limit gives extra relations.
This is due to the specific definition of the loop momenta in string theory.
This approach gives a definite form for the terms that
integrate to zero on the right-hand side of the field-theory relations.
Precise control of the terms that are ambiguous in
field-theoretic constructions will certainly be useful for
deriving the consequences of the monodromy relations at loop
orders in field theory. One other interesting application is the
determination of well adapted classes of momentum-shifting identities 
for integration-by-parts
identities~\cite{Tkachov:1981wb,Chetyrkin:1981qh,Laporta:1996mq,Laporta:2001dd}.
For instance, it was found in~\cite{Bern:2017lpv} that a certain class of
such contributions simplified the study of ultraviolet divergences in
half-maximal supergravity.

Finally, a comment on the co-rank of the momentum kernel. We found it
to be given by $(n-1)!+(n-2)!+2(n-3)!$.  Presumably, these relations
can be interpreted in cohomological terms in the way of
\cite{Mizera:2017cqs}. The counting differs from the number of
solutions to the degenerate forward-tree scattering equations at
one loop, which is
$(n-1)!-2(n-2)!$~\cite{Casali:2014hfa,Geyer:2015bja,Cachazo:2015aol,He:2015yua}
and it would be really interesting to clarify the connection between
the two quantities.

\section*{Acknowledgements}
We would like to thank Simon Badger, Emil Bjerrum-Bohr, Simon Caron-Huot,
John Joseph Carrasco, Michael B. Green, Julio Parra-Martinez, Radu Roiban and Oliver Schlotterer for valuable discussions.

AO is supported in part by the Marie-Curie FP7 grant 631370. The work of PT is supported by STFC grant ST/L000385/1.
The research of PV has received funding the ANR grant reference QST 12 BS05 003 01, and the CNRS grants PICS number 6430. 
PV is partially supported by a fellowship funded by the French
Government at Churchill College, Cambridge. 
PV is grateful to the Mainz Institute for Theoretical Physics (MITP) for its hospitality and its partial support during the completion of this work.
We would like to thanks Claude Duhr and the organisers of the workshop
``LHC and the Standard Model: Physics and Tools'' and the CERN theory
division for the hospitality when completing this work.

\appendix 

\section{Forward limit parametrisation}
\label{sec:forward-limit-param}

If we start with on-shell momenta $p_1+p_2+\dots+p_n=0$,
we may deform the first two of them linearly in complex parameter $z$
in one of the following ways:
\begin{subequations}
\begin{align}
\label{p1p2option1}
 & p_1(z) = \big(\ket{1}+z\ket{\eta_1}\big)[1| , \qquad
   p_2(z) = \big(\ket{2}+z\ket{\eta_2}\big)[2| , \\
\label{p1p2option2}
 & p_1(z) = \big(\ket{1}+z\ket{\eta_1}\big)[1| , \qquad
   p_2(z) = \ket{2}\big([2|+z[\tilde{\eta}_2|\big) ,
\end{align}
\label{p1p2options}%
\end{subequations}
such that $p_i^2(z)=0$ are preserved. We thus obtain
$p_1(z)+p_2(z)+p_3+\dots+p_n \equiv - z q $
for either
$q =\!-\ket{\eta_1}[1|-\ket{\eta_2}[2|$~or~$q =\!-\ket{\eta_1}[1|-\ket{2}[\tilde{\eta}_2|$.
The next step is to define on-shell momenta
$q_2=p_{n+1}(z)$ and $q_1=p_{n+2}(z)$
linearly dependent on $z$, such that
\be
   p_{n+1}(z) + p_{n+2}(z) = z q , \qquad
   p_{n+1}(z=0) = -\ell , \qquad
   p_{n+2}(z=0) = \ell .
\ee
We do this using the following generic solution for the Weyl spinors:
\begin{align}
\label{p5solution}
 & \la_{\alpha}^{n+1}(z)
    = \begin{pmatrix}
         \dfrac{- a \big(q^1-iq^2 + b (q^0-q^3)\big) + z (q \cdot q)}
              {q^0+q^3 + b (q^1+iq^2)} \\
         a
      \end{pmatrix} , \qquad \quad\!
   \lb_{\dot{\alpha}}^{n+1}(z)
    = \begin{pmatrix} 1 \\ b \end{pmatrix} , \\
 & \la_{\alpha}^{n+2}(z)
    = \begin{pmatrix}
         \dfrac{\big(q^1-iq^2 + b(q^0-q^3)\big)\big(a + z(q^1+iq^2)\big)}
              {q^0+q^3 + b (q^1+iq^2)} \\
         -a - z (q^1+iq^2)
      \end{pmatrix} , \quad
   \lb_{\dot{\alpha}}^{n+2}(z)
    = \begin{pmatrix}
         1 \\
         \dfrac{ab - z(q^0+q^3)}{a + z(q^1+iq^2)}
      \end{pmatrix} . \nn
\end{align}

\section{One-loop monodromies in open string}
\label{sec:one-loop-monodromies}

In this appendix we specify and develop some aspects
of the open-string monodromies of~\cite{Tourkine:2016bak}.
We discuss how at one loop the original derivation~\cite{Tourkine:2016bak}
is related to the approach of the subsequent work~\cite{Hohenegger:2017kqy}.

\subsection{Prescription for monodromies and complex logarithm}
\label{sec:comp-recent-string}

\begin{figure}[t]
   \centering
   \input{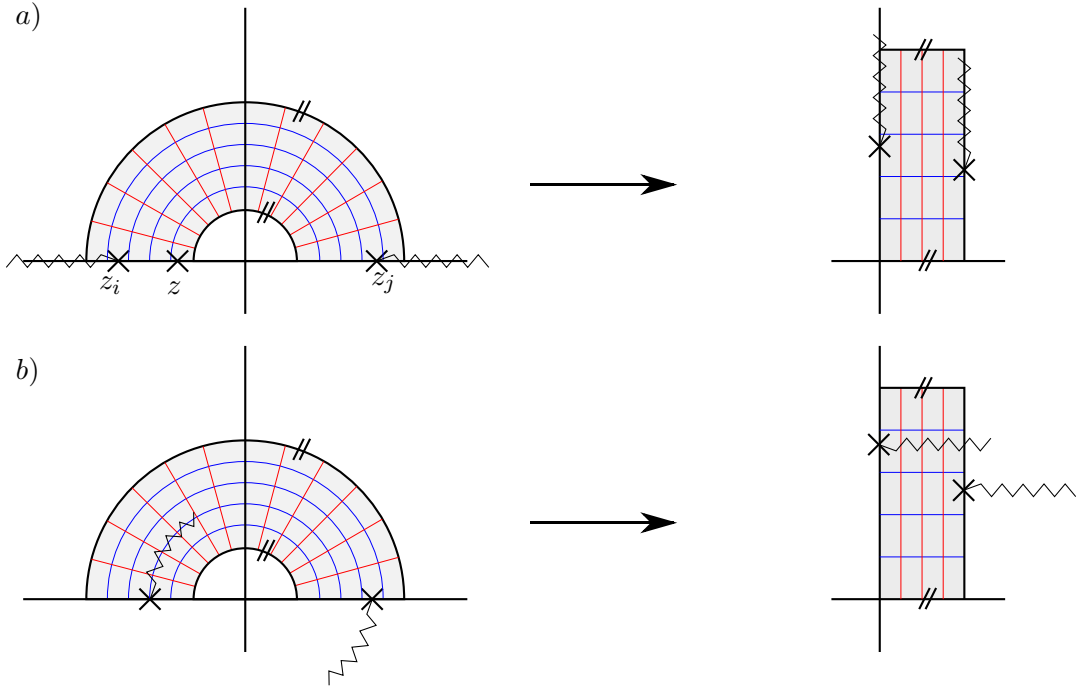}
   \vspace{-7pt}
   \caption{Determination of the complex logarithm on the annulus.
   The choice in~\cite{Tourkine:2016bak} is $(a)$,
   whereas the convention of~\cite{Hohenegger:2017kqy} is $(b)$.}
\label{fig:determination}
\end{figure}

The most notable difference between the approaches
of~\cite{Tourkine:2016bak} and~\cite{Hohenegger:2017kqy} is the choice
of branch cuts for the Green's function.
In~\cite{Tourkine:2016bak} the branch cuts follow the boundary of the annulus and never cross its interior,
as depicted in \fig{fig:determination}$(a)$.
This is an immediate generalisation of the choice
made for the tree-level monodromy relations~\cite{BjerrumBohr:2009rd,BjerrumBohr:2010zs,Stieberger:2009hq}.
In~\cite{Hohenegger:2017kqy} the cuts are chosen to go through the
worldsheet in a maximal number of ways (see \fig{fig:determination}$(b)$), thus requiring to deal with
a variety of additional small contours of integration.
Moreover, the definition of the logarithm in~\cite{Tourkine:2016bak}
does not require the introduction of
bulk terms in the absence of closed-string operator insertions,
to the contrary of the prescription used in~\cite{Hohenegger:2017kqy}. 
Most importantly, converting one prescription to another leads to identical results. 

\subsection{String theory one-loop monodromies at  order $\alpha'$}

To illustrate the general picture given above,
we now expand the four-point relation given in~\cite{Tourkine:2016bak}
to the first order in $\alpha'$
and find that the monodromy relations are satisfied.
The presented analysis is rather similar
to the check of the monodromy relation studied in~\cite{Hohenegger:2017kqy}.

\subsubsection{Planar four-point relation}
\label{sec:planar-contribution}

First, we check the validity of the planar four-point relation
\begin{equation}\label{e:Rel4pt}
   \mathcal  A(1,2,3,4)
 + e^{i\pi\alpha'p_1\cdot p_2}\mathcal A(2,1,3,4)
 + e^{i\pi\alpha'p_1\cdot(p_2+p_3)}\mathcal A(2,3,1,4)
 + \mathcal A(2,3,4|1)[e^{-i\pi\alpha'\ell\cdot p_1}]=0 .
\end{equation}
The planar amplitudes for four gauge bosons are given by 
\begin{multline}
    \mathcal  A(\sigma(1,2,3,4))=t_8F^4\, \int d^{D}\ell\int_0^\infty dt
    \int_{\Delta_{\sigma(1,2,3,4)}} d^3\nu \cr
 \times   e^{-\pi\alpha't\ell^2-2i\pi\alpha' \ell\cdot \sum_{i=1}^4
      p_i\,\nu_i}\prod_{1\leq r<s\leq4}e^{-\alpha' p_r\cdot p_s G(\nu_r-\nu_s)} .
\end{multline}
The one-loop amplitude is proportional to the colour-ordered tree amplitude $t_8F^4$, where the $t_8$-tensor is defined in~\cite[appendix~9.A]{Green:1987mn} and the Green's function is defined
in~\eqn{eq:G-def}.\footnote{The integral is actually divergent from the $t\sim0$ corresponding to
the ultraviolet divergences of the planar open string graph. The
one-loop open string amplitudes are finite only after the addition of
the non-planar Moebius and Klein bottle. The amplitude relations are
valid for any values of $t$ and are not affected by the ultraviolet
behaviour of the individual graphs.}
To perform the $\alpha'$-expansion, we integrate over the loop momentum
and set to $D=10$ the critical dimension for open superstring theory\footnote{We have used the identity
  \begin{equation}
    \sum_{1\leq r<s\leq n} p_r\cdot p_s\, (\Imm(\nu_r-\nu_s))^2= -
    \sum_{r,s=1}^n p_r \cdot p_s\,     \Imm(\nu_r)\,\Imm(\nu_s)
=-\left(\sum_{r=1}^n p_r \,     \Imm(\nu_r)\right)^2
  \end{equation} 
valid for $\sum_{r=1}^n p_r=0$.
Notice that this identity does not require that $p_r^2=0$. 
}
\begin{equation}
      \mathcal  A(\sigma(1,2,3,4))=t_8F^4\,
      \int_0^\infty {dt\over (\alpha' t)^5}
  \int_{\Delta_{\sigma(1,2,3,4)}} d^3\nu\,
    \prod_{1\leq r<s\leq4}e^{-\alpha' p_r\cdot p_s
      (G(\nu_r-\nu_s)+{\pi (\Imm(\nu_r-\nu_s))^2\over t})} ,
\end{equation}
where the colour-ordered tree amplitude $A^{\rm tree}(2,3,4|1)=t_8F^4$.
The non-planar amplitude has a similar expression 
\begin{multline}
   \mathcal  A(2,3,4|1)[e^{-i\pi\alpha'\ell\cdot p_1}]=-t_8F^4\,
      \int\!\dd^{D}\ell \int_0^\infty\!\!\!dt
      \int_{\Delta_{2,3,4|1}}\!\!\!d^3\nu \cr
\times    e^{-\pi\alpha't\ell^2-2i\pi\alpha' \ell\cdot \sum_{i=1}^4
      p_i\,\nu_i-i\pi\alpha' \ell\cdot p_1}
\prod_{1\leq r<s\leq4}e^{-\alpha' p_r\cdot p_s G(\nu_r-\nu_s)} ,
\end{multline}
with $\Ree(\nu_1)=\frac12$ and $\Ree(\nu_a)=0$ with $a=2,3,4$. The overall sign arises from the orientation of boundary on which the
vertex operator 1 is integrated.
Integrating the loop momentum leads to
\begin{multline}
    \mathcal  A(2,3,4|1)[e^{-i\pi\alpha'\ell\cdot p_1}]=-t_8F^4\, \int_0^\infty {dt\over(\alpha' t)^5}
   \int_{\Delta_{2,3,4|1}} d^3\nu \cr
\times     e^{{i\pi\alpha' \over t}p_1\cdot \sum_{i=1}^4 p_i\,\nu_i}
\prod_{1\leq r<s\leq4}e^{-\alpha' p_r\cdot p_s (G(\nu_r-\nu_s) +{\pi(\Imm(\nu_r-\nu_s))^2\over t})} .
\end{multline}
The integration over $\Delta_{2,3,4|1}$ is defined as
$0\leq \Imm(\nu_2) \leq  \Imm(\nu_3) \leq \Imm(\nu_4) =it$ with
$\Ree(\nu_i)=0$ for $i=2,3,4$ and $\Ree(\nu_1)=\frac12$.
Since
\begin{equation}\label{e:t2t1}
   G\left(\frac12+\nu\right)= -\log{\vartheta_2(\nu|it)\over
   \vartheta_1'(0)}=G(\nu)+ \Delta G(\nu) ,
\end{equation}
where we have introduced 
\begin{equation}\label{e:DG}
\Delta G(\nu)=-\log\cot(\pi
    \nu)-4\sum_{m\geq0} {q^{2m+1}\over 1-q^{2m+1}} \, {\cos(2\pi (2m+1)
    \nu)\over (2m+1)} .
\end{equation}
Performing the expansion of~(\ref{e:Rel4pt}) at the first order in
$\alpha'$, as a function of the proper time $t$,
its integrand reads
\begin{multline}\label{e:Identity}
 i\pi\alpha'\left(\int_{    \Delta_{2,1,3,4}}d^3\nu\, p_1\cdot p_2
-\int_{\Delta_{2,3,1,4}}d^3\nu \,  p_1\cdot p_4\right)
- \int_{\Delta_{2,3,4|1}}d^3 \nu
     {\pi\alpha'p_1\cdot \sum_{i=1}^4 p_i\,\nu_i\over t}\cr
 +\int_{\Delta_{1,2,3,4}\cup
    \Delta_{2,1,3,4}\cup\Delta_{2,3,1,4}}d^3\nu\mathcal Q - \mathcal Q\stackrel{?}{=} 0 ,
\end{multline}
where we introduced the short-hand notation
\begin{equation}
\label{e:cQ}
  \mathcal Q= \sum_{1\leq r<s\leq 4} p_r\cdot p_s \,
  \left(G(\nu_r-\nu_s)+{\pi (\Imm(\nu_r-\nu_s))^2\over t}\right) .
\end{equation}
Using that a consequence of~\eqref{e:t2t1}  allows to re-express the
non-planar contribution as a sum of planar contributions
\begin{equation}\label{e:NPtoP}
  \int_{\Delta_{2,3,4|1}}d^3 \nu\,\mathcal Q=\int_{\Delta_{1,2,3,4}\cup
    \Delta_{2,1,3,4}\cup\Delta_{2,3,1,4}}d^3\nu \, \left(\mathcal Q -\Delta
  \mathcal Q\right) ,
\end{equation}
where
\begin{equation}\label{e:DcQ}
  \Delta \mathcal Q= p_1\cdot p_2 \Delta G(\nu_1-\nu_2)+p_1\cdot p_3 \Delta G(\nu_1-\nu_3)+p_1\cdot p_4 \Delta G(\nu_1-\nu_4) .
\end{equation}
The contributions from the phases compensate exactly each other
because
\begin{equation}
     -\int_{\Delta_{2,3,4|1}}d^3\nu
     {\pi\alpha' \over t}p_1\cdot \sum_{i=1}^4 p_i\nu_i= -{\pi\alpha'
        t^3\over
      6}(p_1\cdot p_2-p_1\cdot p_4) 
\end{equation}
is the opposite of
\begin{equation}
  i\pi \alpha' \left(\int_{\Delta_{2,1,3,4}} d^3\nu\, p_1\cdot p_2 - \int_{\Delta_{2,3,1,4}} d^3\nu\, p_1\cdot p_4 \right)={\pi\alpha' t^3\over
  6} (p_1\cdot p_2-p_1\cdot p_4) . 
\end{equation}
After cancellation of the integrals of $\mathcal Q$ using \eqn{e:NPtoP},
the relation~\eqref{e:Identity} to prove becomes
\begin{equation}\label{e:Identity2}
\int_{\Delta_{2,3,4|1}}d^3 \nu  \Delta \mathcal Q\stackrel{?}{=} 0 .
\end{equation}

We compute the integral by first evaluating the contributions
to the $q$-expansion of $\Delta \mathcal Q$ from $q$-expansion in
$\Delta G(\nu)$ in \eqn{e:DG}
We find that setting $q=\exp(-2\pi t)$
\beal
    \int_{\Delta_{2,3,4|1}}d^3 \nu  \Delta \mathcal Q\Big|^{q\text{-exp}}
    = i\,p_1\cdot p_4\,
    \bigg\{ \frac{(\log q)^2}{64\pi}
          + \frac{7(\log q)\zeta(3)}{32\pi^3} & \\
          + \frac{(\log q)^2 (\textrm{Li}_2(q)-\textrm{Li}_2(-q))}{16 \pi^3
}-\frac{ \log q\,(\textrm{Li}_3(q)-\textrm{Li}_3(-q))}{8 \pi^3} & \bigg\} .
\eeal

To evaluate the zero mode, we have made use of the program
{\tt HyperInt} by Erik Panzer~\cite{Panzer:2014caa} to find
\begin{subequations}
\begin{align}
   \int_{\Delta_{2,3,4|1}} d^3\nu \log(\cot\pi(\nu_1-\nu_2)) &=
   \frac{\log (q) \left(4\text{Li}_3\left(-q\right)-4 \text{Li}_3(q)+\pi ^2 \log (q)+7 \zeta(3)\right)}{32 \pi ^3\,i} , \\
   \int_{\Delta_{2,3,4|1}} d^3\nu \log(\cot\pi(\nu_1-\nu_3)) &=
   \frac{\log (q) \left(4 \text{Li}_3\left(-q\right)-4 \text{Li}_3(q)+ \pi ^2 \log (q)+7
   \zeta (3)\right)}{32 \pi ^3\,i} , \\
   \int_{\Delta_{2,3,4|1}} d^3\nu \log(\cot\pi(\nu_4-\nu_1)) &=
   \frac{\left(4 \text{Li}_2(-q)-4 \text{Li}_2(q)+\pi ^2\right) \log ^2(q)}{64 \pi ^3\,i} .
\end{align}
\end{subequations}

Summing over these contributions leads to complete cancellation
between the integral over the zero-mode part
and the $q$-expansion in $\Delta \mathcal Q$, establishing that for
all values of the proper time $t$
\begin{equation}\label{e:Identity3}
\int_{\Delta_{2,3,4|1}}d^3 \nu  \Delta \mathcal Q\stackrel{\checkmark}{=} 0.
\end{equation}
This completes the verification of the amplitude relation~\eqref{e:Rel4pt} to the first order in the $\alpha'$-expansion.

Notice that the vanishing is satisfied for all values of the proper
time $t$, and therefore there is no need of considering a
regularisation of the ultraviolet divergence (for $t\sim 0$) of the
annulus graph.

\subsubsection{Non-planar four-point relation}
\label{sec:non-planar-four}

The non-planar $n$-point
monodromy relation is given by\footnote{We correct a sign mistake
in~\cite{Tourkine:2016bak} for the non-planar phases.
The non-planar cuts were incorrectly determined to be downward cuts,
while in fact they are upward cuts,
as pictured in the right column of \fig{fig:determination}$(a)$.}
\beal
  \label{eq:monod-n-generic-correct}
  \cA(1,2,\ldots ,p|p+1, \ldots ,n) 
 +
\sum_{i=2}^{p-1} e^{i\alpha'\pi p_1\cdot p_{2\cdots i} }
  \cA(2,\ldots ,i,1,i+1,\ldots,p |p+1,\ldots ,n) & \\
 -\sum_{i=p}^n \big(  e^{-i\alpha'\pi p_1\cdot p_{i+1\cdots n}}\times
  \cA(2,\ldots ,p|p+1 ,\ldots ,i,1,i+1,\ldots ,n)
  [e^{-i\pi\alpha'\ell\cdot p_1}]\big) &=0 . \!\!
\eeal
We now wish to verify the following non-planar four-point relation: 
\beal\label{e:Rel4ptNP}
    \mathcal A(1,2|3,4)
  + e^{i\pi\alpha'p_1\cdot (p_3+p_4)}
    \mathcal A(2|1,3,4) [e^{-i\pi\alpha'\ell\cdot p_1}] & \\
  + e^{i\pi\alpha'p_1\cdot p_4}
    \mathcal A(2|3,1,4) [e^{-i\pi\alpha'\ell\cdot p_1}] &
  + \mathcal A(2|3,4,1)[e^{-i\pi\alpha'\ell\cdot p_1}] = 0 .
\eeal
The first non-planar amplitude is given by
\begin{equation}
   \mathcal  A(1,2|3,4)= t_8F^4\,
      \int\!\dd^{D}\ell \int_0^\infty\!\!\!dt
      \int_{\Delta_{1,2|3,4}} \!\!\!\!d^3\nu 
  e^{-\pi\alpha't\ell^2-2i\pi\alpha' \ell\cdot \sum_{i=1}^4
      p_i\,\nu_i}
\prod_{1\leq r<s\leq4}e^{-\alpha' p_r\cdot p_s G(\nu_r-\nu_s)} ,
\end{equation}
where $\Delta_{12|34}$ is defined by $\nu_2=it$,
$0\leq\Imm(\nu_1)\leq t$ and $0\leq \Imm(\nu_4)\leq \Imm(\nu_3)\leq t$.
We integrate over the loop momentum to get the expression
\begin{equation}
    \mathcal  A(1,2|3,4)= t_8F^4\, \int_0^\infty {dt\over (\alpha't)^5}
     \int_{\Delta_{1,2|3,4}} d^3\nu 
     \,
\prod_{1\leq r<s\leq4}e^{-\alpha' p_r\cdot p_s
  (G(\nu_r-\nu_s)+{\pi(\Imm(\nu_r-\nu_s))^2\over t})} .
\end{equation}
The next amplitude is
\begin{multline}
   \mathcal  A(2|1,3,4)[e^{-i\pi\alpha'\ell\cdot p_1}]=-t_8F^4\,
      \int\!\dd^{D}\ell \int_0^\infty\!\!\!dt
      \int_{\Delta_{2|1,3,4}} d^3\nu \,  e^{-\pi\alpha't\ell^2-2i\pi\alpha' \ell\cdot \sum_{i=1}^4
      p_i\,\nu_i-\pi\alpha' \ell\cdot p_1}\cr
\times
\prod_{1\leq r<s\leq4}e^{-\alpha' p_r\cdot p_s G(\nu_r-\nu_s)} ,
\end{multline}
where the overall sign comes from the contour of integration, and
$\Delta_{2|1,3,4}$ is defined by $\nu_2=it$ and
$0\leq \Imm(\nu_4)\leq \Imm(\nu_3)\leq \Imm( \nu_1)\leq t$, with $\Ree(\nu_2)=0$ and
$\Ree(\nu_i)=\frac12$ for $i=1,3,4$.
Equivalent definitions determine the amplitudes $\mathcal
A(2|3,1,4)[e^{-i\pi\alpha'\ell\cdot p_1}]$ and $\mathcal  A(2|3,4,1)[e^{-i\pi\alpha'\ell\cdot p_1}]$.
Integrating over the loop momentum gives
\begin{multline}
    \mathcal  A(2|1,3,4)[e^{-i\pi\alpha'\ell\cdot p_1}]=-t_8F^4\,
     \int_0^\infty {dt\over (\alpha't)^5}
     \int_{\Delta_{2|1,3,4}} d^3\nu   e^{{i\pi\alpha'\over t} p_1\cdot
       \sum_{i=1}^4 p_i \,\nu_i}\cr
\times  
\prod_{1\leq r<s\leq4}e^{-\alpha' p_r\cdot p_s
  (G(\nu_r-\nu_s)+{\pi(\Imm(\nu_r-\nu_s))^2\over t})} .
\end{multline}

Like before, integrating out the loop momentum results in phase
factors that can be explicitly computed at first order in $\alpha'$
\begin{multline}
\int_{\Delta_{2|1,3,4}\cup \Delta_{2|3,1,4}\cup\Delta_{2|3,4,1}} \!\!\!\!\!d^3\nu
  {\pi\alpha' p_1\cdot \sum_{i=1}^4 p_i\,\nu_i\over t} 
+i\pi\alpha' \left(  \int_{\Delta_{2|1,3,4}}d^3\nu\, p_1\cdot p_{34}+
  \int_{\Delta_{2|3,4,1}}d^3\nu  \,p_1\cdot p_4\right)\cr+
   \int_{\Delta_{12|34}}d^3\nu \mathcal Q- \int_{\Delta_{2|1,3,4}\cup \Delta_{2|3,1,4}\cup\Delta_{2|3,4,1}}d^3\nu \mathcal Q \stackrel{?}{=}0 ,
\end{multline}
where $\mathcal Q$ has been introduced in \eqn{e:cQ}.
We can rewrite this integral as
\begin{multline}\label{e:Id2}
\int_{\Delta_{2|1,3,4}\cup \Delta_{2|3,1,4}\cup\Delta_{2|3,4,1}} \!\!\!\!\!d^3\nu
  {\pi\alpha'p_1\cdot \sum_{i=1}^4 p_i\,\nu_i\over t} 
+i\pi\alpha' \left(\int_{\Delta_{2|1,3,4}}d^3\nu \,p_1\cdot p_{34}
+\int_{\Delta_{2|3,1,4}}d^3\nu \,p_1\cdot p_{4}\right)\cr
- \int_{\Delta_{12|34}}d^3\nu \Delta\mathcal Q \stackrel{?}{=}0 ,
\end{multline}
where $\Delta\mathcal Q$ is defined in~\eqn{e:DcQ}.
We then find
\begin{subequations}
\begin{align}
     \int_{\Delta_{2|1,3,4}\cup \Delta_{2|3,1,4}\cup\Delta_{2|3,4,1}} d^3\nu
     {\pi\alpha' \over t}p_1\cdot \sum_{i=1}^4 p_i\,\nu_i=& {\pi\alpha'
        t^3\over
      6}(p_1\cdot p_2-p_1\cdot p_4) ,\\
\int_{\Delta_{2|1,3,4}}d^3\nu (i\pi\alpha' p_1\cdot p_{34})&=-{\pi\alpha' t^3\over
                                                       6} p_1\cdot p_2 ,\\
  \int_{\Delta_{2|3,1,4}}d^3\nu  (i\pi\alpha' p_1\cdot p_4)&={\pi\alpha' t^3\over
                                                       6} p_1\cdot p_4 ,
\end{align}
\end{subequations}
therefore we see cancellation of the phase factors in the first line of \eqn{e:Id2}.
For the second line we find for the $q$-expansion of $\Delta \mathcal Q$
\begin{multline}
 \int_{\Delta_{1,2|3,4}} d^3\nu \Delta\mathcal Q|^{q-exp}= -ip_1\cdot p_2\,
 {\log(q)^2\over 64\pi^3} \, \left( {3\over2}\zeta(2) +
   \textrm{Li}_2(q)-\textrm{Li}_2(-q)\right)\cr
-i p_1\cdot p_2 \,{\log(q)\over 64\pi^3}\, \left(-{7\over 4} \zeta(3)+\textrm{Li}_3(q)-\textrm{Li}_3(-q)\right) ,
\end{multline}
which is cancelled against the constant terms from $\Delta \mathcal Q$.
Hence we arrive at
\begin{equation}
   \int_{\Delta_{1,2|3,4}}\!\!\!d^3\nu\,\Delta\mathcal Q
   \stackrel{\checkmark}{=}0 ,
\end{equation}
which proves the identity~\eqref{e:Id2}.

\bibliographystyle{JHEP}
\bibliography{biblio}
\end{document}